\documentclass[a4paper,12pt,english]{article}
\usepackage[english]{babel}
\usepackage{amsfonts}
\usepackage{amsmath}
\usepackage{graphicx}
\usepackage{subfigure}

\usepackage{array}
\begin{document}

\title{Wicking in a powder}
\author{P.S. Raux\footnote{PMMH, UMR 7636 du CNRS, ESPCI - 75005 Paris, France and LadHyX, UMR 7646 du CNRS, Ecole Polytechnique -91128 Palaiseau Cedex, France} , H. Cockenpot$^*$, M. Ramaioli\footnote{Department Food Science \& Technology,
Nestlé Research Center - 1000 Lausanne 26, Switzerland}, D. Qu\'er\'e$^*$ and C. Clanet$^*$}

\date{}
\maketitle

\begin{abstract}
We investigate the wicking in granular media by considering layers of grains at the surface of a liquid, and discuss the critical contact angle below which spontaneous impregnation takes place. This angle is found to be on the order of $55^\circ$ for monodisperse layers, significantly smaller than $90^\circ$, the threshold value for penetrating assemblies of tubes: owing to geometry, impregnating grains is more demanding than impregnating tubes. We also consider the additional effects of polydispersity and pressure on this wetting transition, and discuss the corresponding shift observed on the critical contact angle.
\end{abstract}

\section{Introduction}

\begin{figure}[h]
\begin{center}
\includegraphics[width=0.8\textwidth]{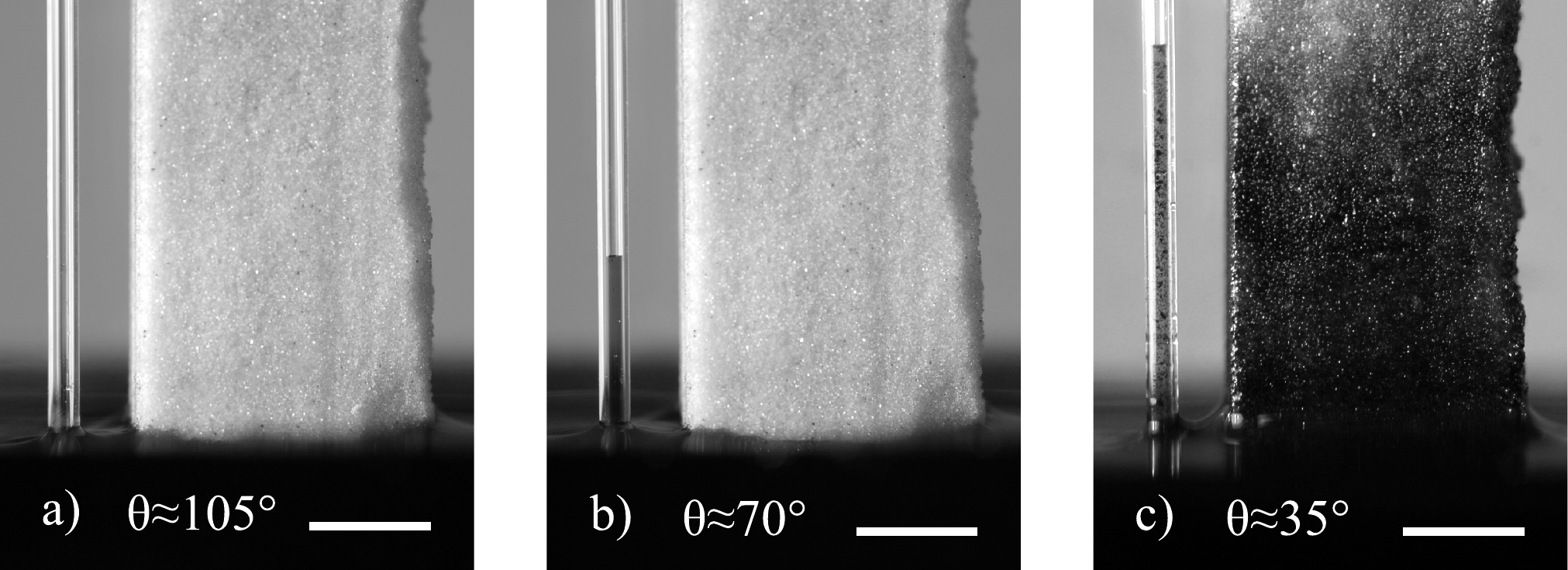}
\caption{A silanized capillary tube and a porous media made of silanized glass beads are put in contact with various water-ethanol baths, changing the contact angle $\theta$. (a) No wicking is observed at high contact angles. (b) Wicking in the tube occurs when the contact angle is below $90^\circ$  but the grains can remain dry. (c) Spontaneous impregnation in the porous media is observed only below a critical contact angle $\theta^\star$ significantly smaller than $90^\circ$. Scale bars correspond to 5~mm.}
\end{center}
\label{fig:intro}
\end{figure}



Spontaneous invasion of pores by liquids is driven by a reduction of surface energy. This phenomenon is relevant to many natural applications such as soil imbibition by rain \cite{Green1911studies,doerr2000soil,doerr2006occurrence}, soil decontamination \cite{amro2004factors} or plant physiology \cite{taiz1998plant,ponomarenko2011universal}. It is also of broad industrial relevance for oil extraction, civil engineering \cite{hall2002water}, absorbent consumer goods \cite{staples2002wicking,brielles2008dissolution}, paper and textile design \cite{bico2007precursors,alava2006physics}, chromatographic \cite{wall2005thin} and microfluidic processes \cite{bouaidat2005surface}. An important application in food industry is the reconstitution of a beverage after wetting of a dehydrated powder \cite{forny2011wetting}. Depending on the field of interest, impregnation can either be beneficial (food industry), or detrimental (civil engineering).

 Figure \ref{fig:intro} compares wicking in porous media made of spherical beads and in a capillary tube. For large contact angles, there is no wicking (Figure \ref{fig:intro}a). Impregnation only occurs if the contact angle is below a critical value: ninety degrees for the capillary tube (Figure \ref{fig:intro}b), and a much lower one for the beads (Figure \ref{fig:intro}c). This experiment emphasizes the role of geometry in wicking: a porous medium made of grains or beads substantially differs from cylindrical capillaries, because cross sectional area and wall orientation vary along the pores. This modifies the condition of wicking, as also shown for other special porous media, such as wedges \cite{concus1969behavior} and micro-textures on solid surfaces\cite{dG2003capillarity,blow2009imbibition}. However, the classical theories  for cylindrical tubes \cite{hauksbee1708several,jurin1717account,laplace1805theory} are still broadly applied to powders.

 In the general case of a compact pile of grains, the maximum penetration height can be computed by appropriately adapting Jurin's theory \cite{dG2003capillarity}. In the same spirit, the classical kinetics of Lucas-Washburn \cite{lucas1918ueber,washburn1921dynamics} can be modified by considering the effect of the variable cross section of the granular pores on the viscous dissipation, which generates the rich impregnation kinetics of layered beads  \cite{fries2008analytic,reyssat2009imbibition}. In these works, the capillary driving force is often considered as constant along pore length. Reyssat \textit{et al.} \cite{reyssat2009imbibition} considered for instance complete wetting and included an adjustable parameter in the capillary term. Fries \textit{et al.} \cite{fries2008analytic} assumed that spontaneous impregnation in granular beds occurs if the contact angle of the liquid with the channels is lower than ninety degrees, that is, the condition for pores having walls parallel to the pore axis.
Tsori \cite{tsori2006discontinuous} and Czachor \cite{czachor2006modelling} theoretically studied the capillary penetration in channels of varying cross section, including sinusoidal capillaries. The former showed the existence of multiple equilibrium positions, while the latter predicted the existence of a critical wetting angle for capillary rise that depends on the sinusoidal wall waviness. Lago \textit{et al.} \cite{lago2001capillary} discussed the optimal angle for the highest capillary rise, and found $\theta=15.4^\circ$ for compact piles of spherical grains. B\'an \textit{et al.} \cite{ban1987condition} and Shirtcliffe \textit{et al.} \cite{shirtcliffe2006critical} investigated spontaneous wicking in such piles. They predicted the existence of a critical contact angle $\theta^\star < 90^\circ$ for wicking in this geometry,  and produced experimental evidences for this result. These works will be further commented in sections 4 and 5.

In this article, we report experiments and discuss a model, showing the existence of an acute contact angle above which the impregnation of a heap of grains can be blocked, as seen in Figure \ref{fig:intro}.
 We also discuss how this critical angle depends on polydispersity, hydrostatic forcing and disorder. These results can be used either to improve or to block capillary invasion, depending on the system of interest.


\section{Experimental set-up and protocol} \label{sect:setup}


\begin{figure}[h!]
\centering
\subfigure[]{\includegraphics[width=0.3\textwidth]{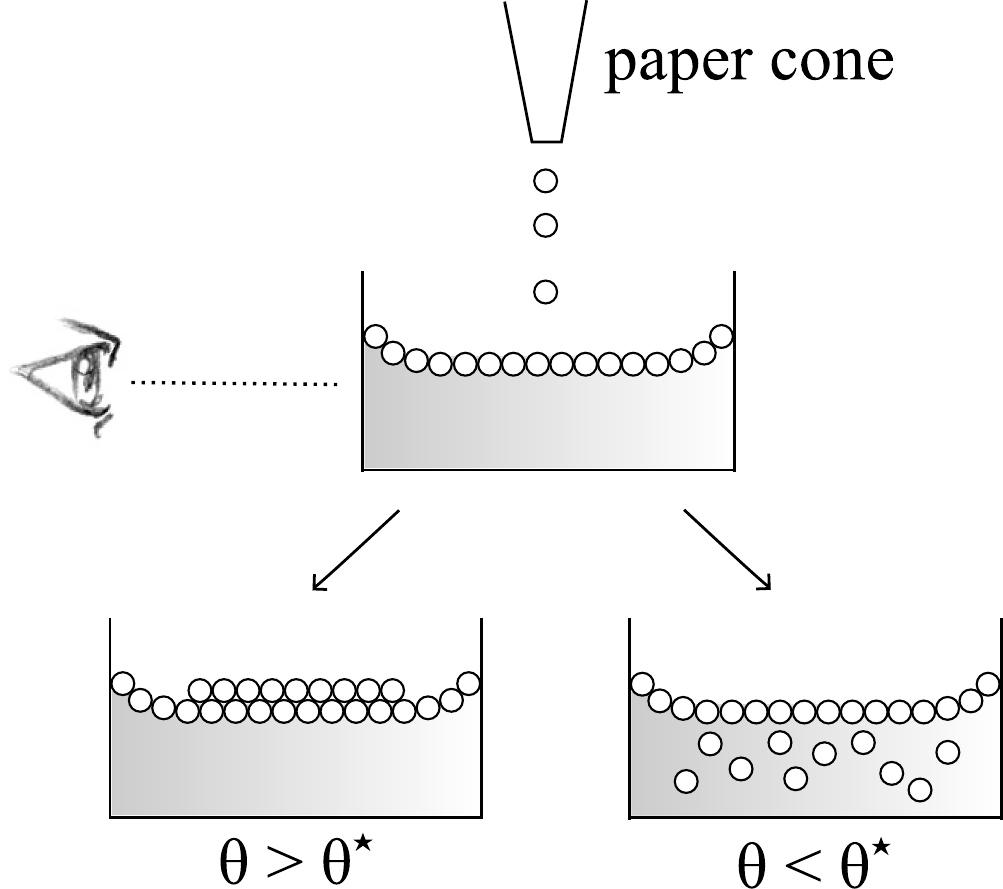}\label{fig:mono-schema}}
\subfigure[]{\includegraphics[width=0.3\textwidth]{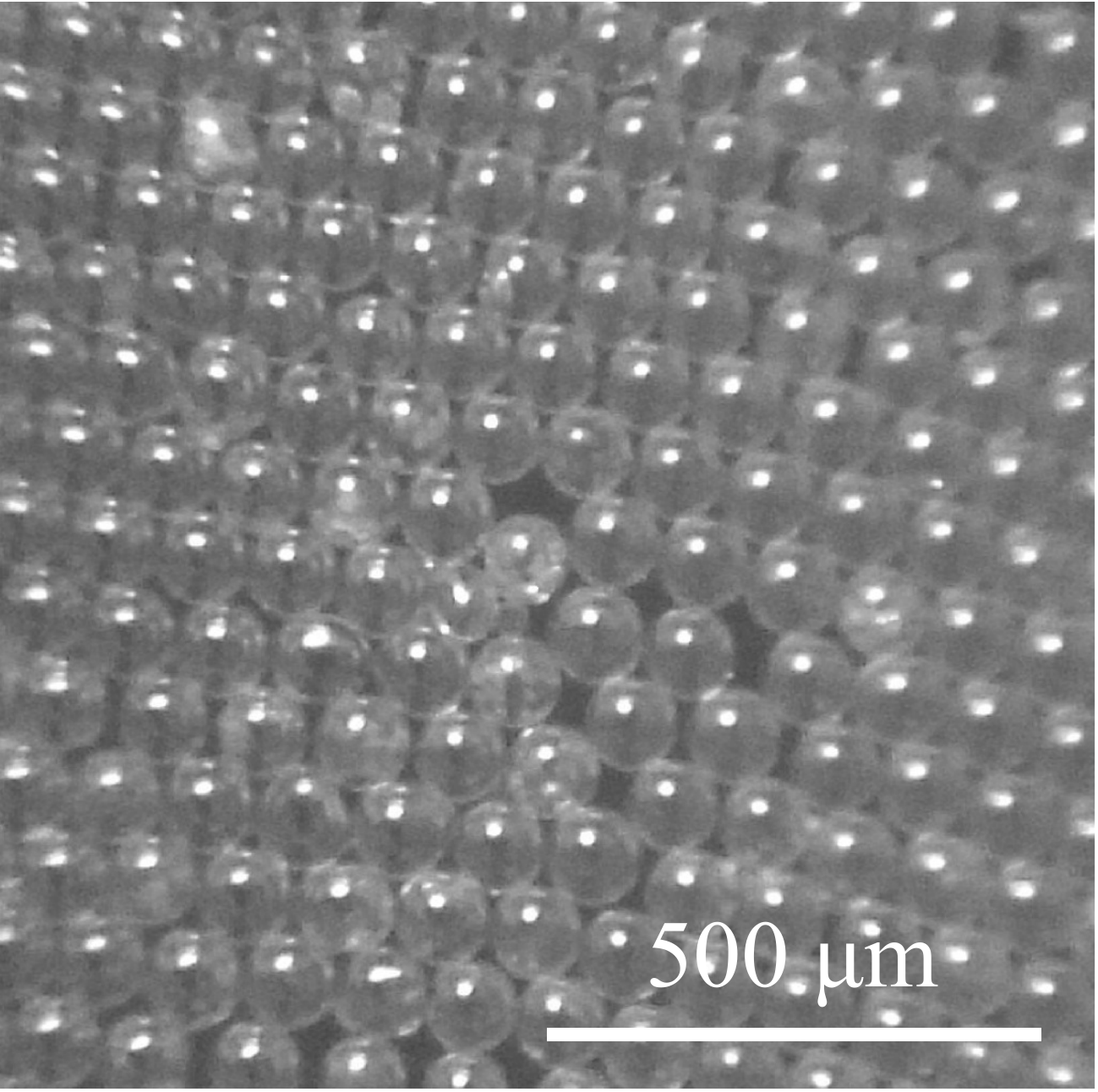}\label{fig:beads}}
\subfigure[]{\includegraphics[width=0.35\textwidth]{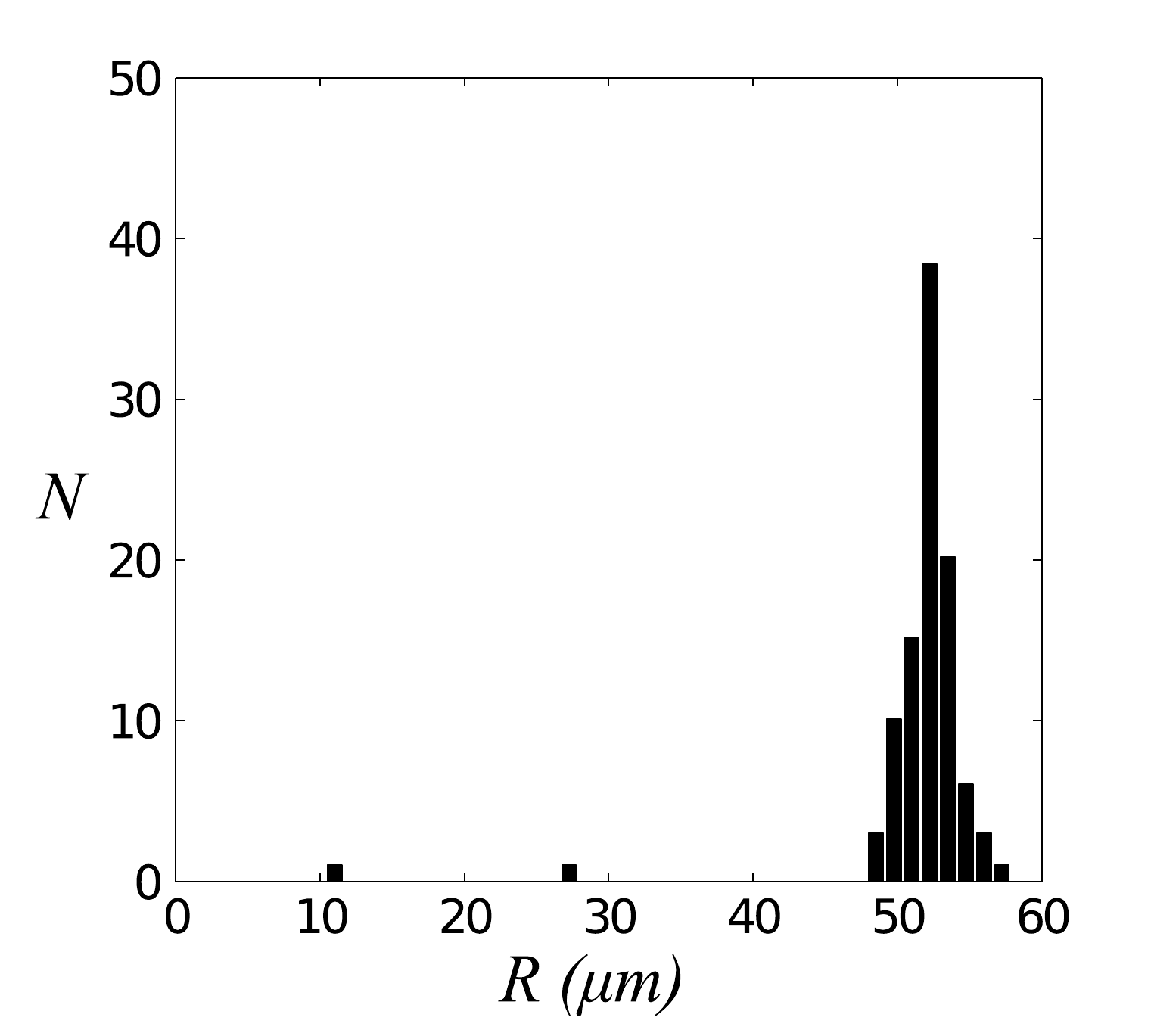}\label{fig:hist}}
\caption{(a) Experimental set-up and sketch of the wicking criterion. (b) Silanized glass beads of radius $R_1=52\pm2~\mu m$ at the surface of a water/ethanol bath. (c) Corresponding distribution of beads radii for a sample of 100 beads.}
\label{fig:setup}
\end{figure}
\begin{figure}[t]
\centering
\includegraphics[width=0.6\textwidth]{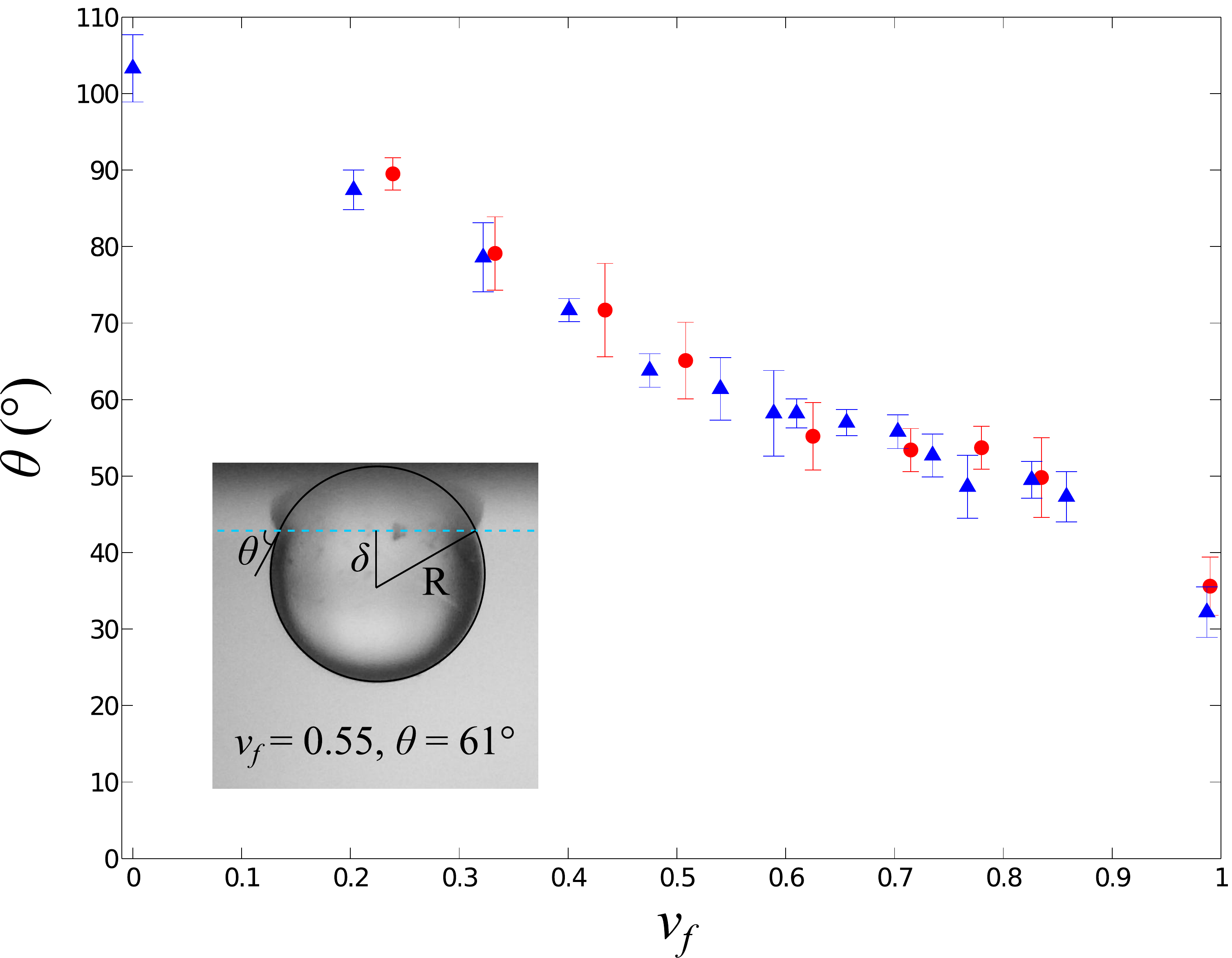}
\caption{Contact angle $\theta$ between silanized glass beads of radius $R=52~\mu m$ (red circles) or $R=256~\mu m$ (blue triangles), and water-ethanol mixtures, as a function of the ethanol volume fraction $v_f$. The contact angle can be continuously varied between $105^\circ \pm 5^\circ$ and $35^\circ \pm 5^\circ$ by increasing the ethanol fraction. The insert shows a typical measurement of the contact angle on a bead (with $R=256~\mu m$ and $v_f = 0.55$). The dashed line indicates the liquid/air interface. }

\label{fig:tvsVf}
\end{figure}

The experiment consists of depositing dry silanized grains of glass at the surface of a liquid bath filling a centimetric transparent cell (Figure \ref{fig:setup}a). The liquid-grain interface is observed with a camera, magnification x2. Our wicking criterion is the detachment of grains from the interface, which then fall in the bath due to a density $\rho_s = 2450~\mathrm{kg/m^3}$ higher than the one of water. Conversely, the liquid does not penetrate the pile if all the grains remain at the surface.

The grains are borosilicate glass beads (from Sigmund Lindner),  sieved in order to reduce the polydispersity (Figures 2b and 2c). Several radii $R$ are used: $25\pm2~\mu m$, $52\pm2~\mu m$, $100\pm5~\mu m$ and $256\pm13~\mu m$. The glass beads, initially hydrophilic, are silanized with 1H-1H-2H-2H perfluoro-octyltriethoxysilane to be hydrophobic, following the protocol for fluorination by Qian \textit{et al.} \cite{qian2005fabrication}.  
The wetting properties of the system are tuned by using mixtures of water and ethanol. The ethanol volume fraction $v_f$ in water is controlled by measuring the density of the mixture, referring to tables in Handbook of Chemistry and Physics \cite{hodgeman1949handbook}. The capillary length $a=\sqrt{\frac{\gamma}{\rho g}}$ (where $\gamma$ and $\rho$ are the surface tension and density of the liquid, and $g$ the gravity constant) varies from $2.7~\mathrm{mm}$ for deionized water to $1.7~\mathrm{mm}$ for pure ethanol, always larger than the beads radius.



In order to measure the contact angle $\theta$, single beads (with $R\ll a$) are placed at the surface of the liquid, and $\theta$ is deduced from the distance $\delta=R \cos \theta$ between the bead center and the liquid interface (insert in Figure \ref{fig:tvsVf}).  Pictures are taken with a Ricoh GX200 camera trough a binocular x20. The position of the liquid-gas interface is determined by the reflection of the beads on the surface. Measurement of contact angle for each $v_f$ is repeated on 10 different beads. The contact angle $\theta$ with deionized water ($v_f=0$) is $105^\circ \pm 5^\circ$, and it falls down to $35^\circ \pm 5^\circ$ for pure ethanol ($v_f=1$). As presented in Figure \ref{fig:tvsVf}, water/ethanol mixtures allow us to obtain intermediate contact angles, which continuously decrease from $105^\circ$ to $35^\circ$ as $v_f$ passes from $0$ to $1$.

\section{Experimental results} \label{sect:expmt}

\begin{figure}[h!]
\centering
\subfigure[$v_f=0.4, \theta=75^\circ \pm 5^\circ$]{\includegraphics[width=0.3\textwidth]{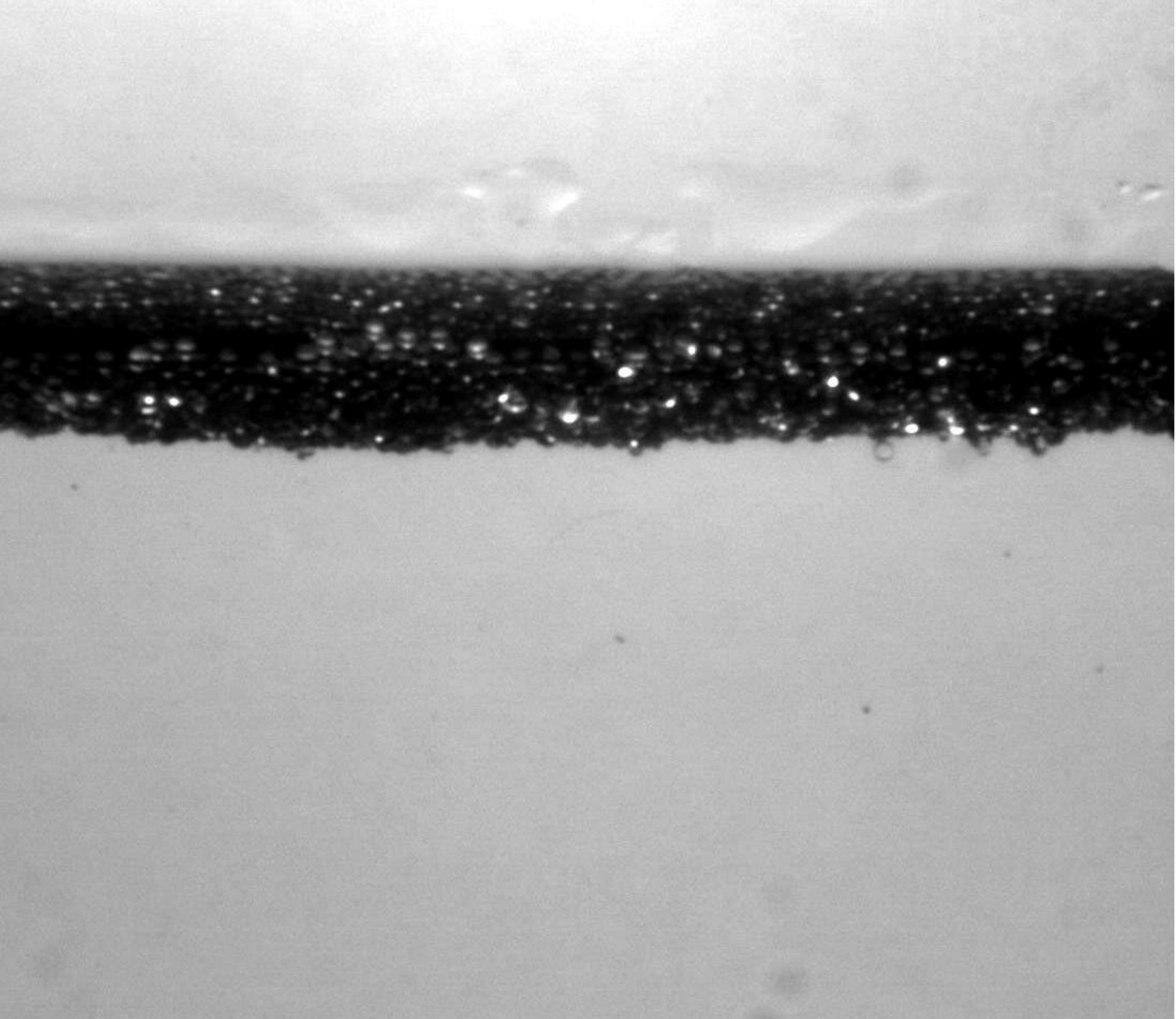}\label{fig:h0ni}}
\subfigure[$v_f=0.65,\theta=58^\circ \pm7^\circ$]{\includegraphics[width=0.3\textwidth]{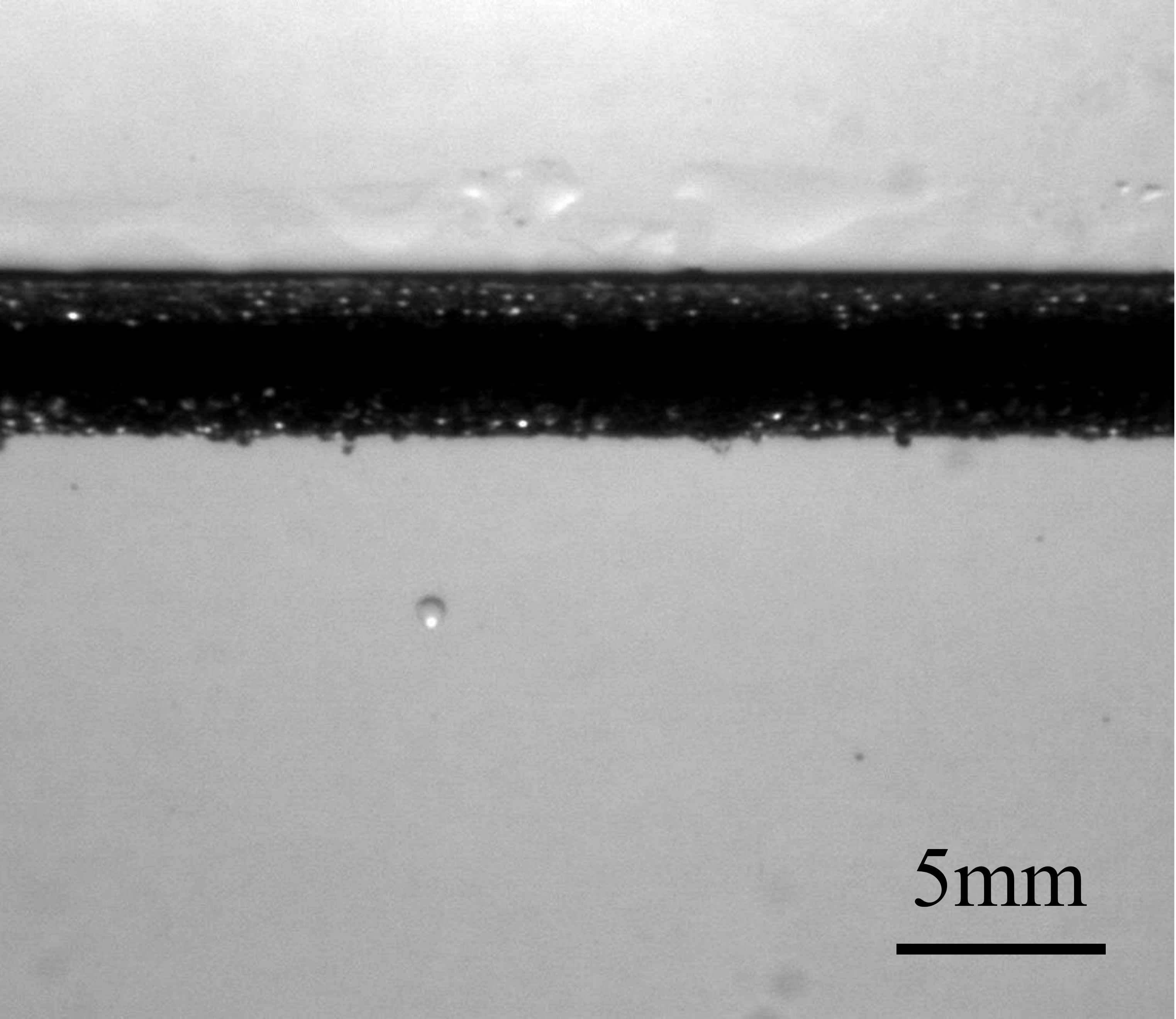}\label{fig:h0lim}}
\subfigure[$v_f=0.7,\theta=56^\circ \pm 5^\circ$]{\includegraphics[width=0.3\textwidth]{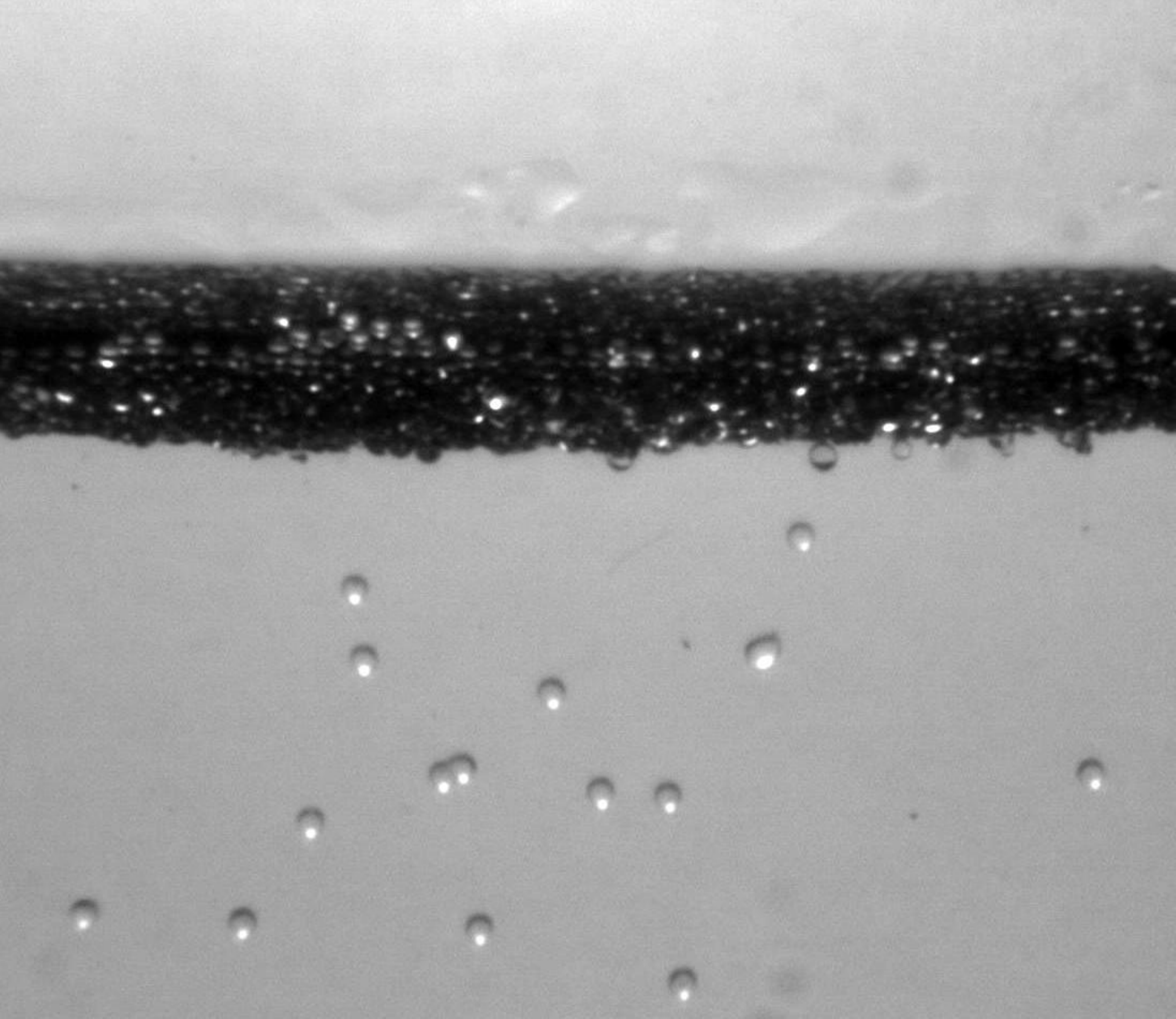}\label{fig:h0i}}
\hspace{.03\textwidth}
\caption{Side-views of our experiment for different ethanol volume fractions $v_f$. A pile of grains is made at the interface between the bath and air. The apparent thickness (in black) of the interface comes from the menisci at the edges. Wicking is deduced from the presence of grains falling in the bath: (a) no impregnation; (b) limit case; (c) impregnation. Glass beads radius here is $256\pm13~\mu m$.} 
\label{fig:impr}
\end{figure}

We deposit a uniform layer of dry grains at the surface of a bath of ethanol fraction $v_f$ (Figure \ref{fig:impr}). A meniscus forms along the cell walls, which generates the dark zone observed in the figure. The actual thickness of the powder only represents a few layers of beads (typically 4). Two regimes are observed. In Figure \ref{fig:impr}a,  no bead falls into the bath, and the pile remains dry. In Figure \ref{fig:impr}c, beads detach and fall as the liquid invades the pile. Figure \ref{fig:impr}b shows the limit of impregnation, where only very few beads fall. The wicking transition occurs for an ethanol fraction $v_f^\star=0.65\pm0.03$ (for $R=256~\mu m$), which corresponds to a contact angle $\theta^\star = 58^\circ\pm7^\circ$. $\theta^\star$ is a critical angle: if we have $\theta<\theta^\star$, wicking is observed; if not, the pile remains dry. The results for different grain radii are presented in Table \ref{tablo}, showing that $\theta^\star$ hardly depends on $R$ for grains smaller than the capillary length ($R<<a$). The highest value of the critical contact angle corresponds to the largest radius. Since the density  $\rho_s$ of the grain is significantly larger than the density of the liquid, gravitational corrections are expected for the experiments with  $R=256~\mu m$.

\renewcommand{\arraystretch}{1.8}
\begin{table}[t]
\centering
\begin{tabular}{|c|c|c|}
\hline
$R\: (\mu m)$ &  $v_f^\star$  & $\theta ^\star \: (^\circ)$\\
\hline
$25 \pm 2$ & $0.73 \pm 0.01$ & $53 \pm 6$\\
\hline
 $52 \pm2$ & $0.73 \pm 0.01 $ &$ 53 \pm 6$ \\
\hline
 $100 \pm 5$&$0.70\pm0.01$&$55\pm6$\\
\hline
\hspace{0.05\textwidth} $256 \pm 13$\hspace{0.05\textwidth}  & \hspace{0.05\textwidth} $0.65 \pm 0.03 $\hspace{0.05\textwidth} & \hspace{0.05\textwidth} $58 \pm7$\hspace{0.05\textwidth} \\
\hline
\end{tabular}

\caption{\label{tablo} Wicking transition in terms of critical ethanol fraction $v_f^\star$, and corresponding contact angle  $\theta^\star$, for different beads of radius $R$ smaller than the capillary length $a$.}
\end{table}

\section{Model} \label{sect:theta0}
\subsection{Interface equilibrium}

\begin{figure}[h]
\centering
\subfigure[]{\includegraphics[width=0.35\textwidth]{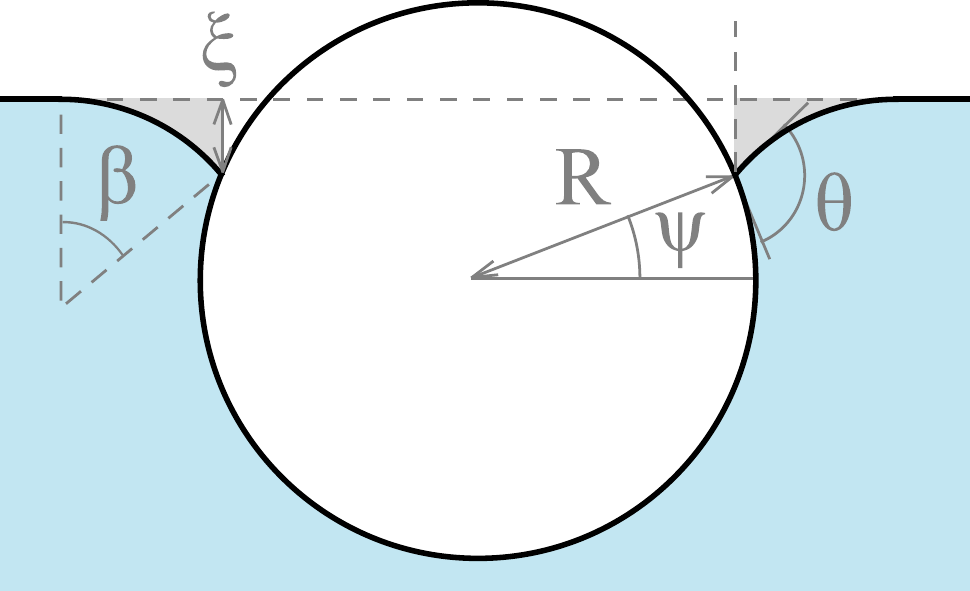}} 
\hspace{.1\textwidth}
\subfigure[]{\includegraphics[width=0.35\textwidth]{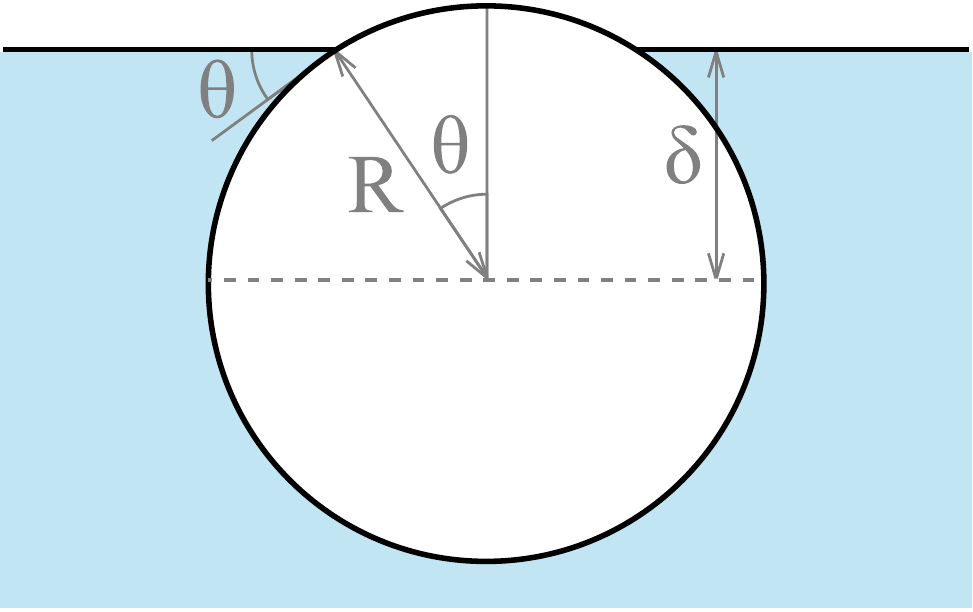}\label{fig:grains1}}
\subfigure[]{\includegraphics[width=0.4\textwidth]{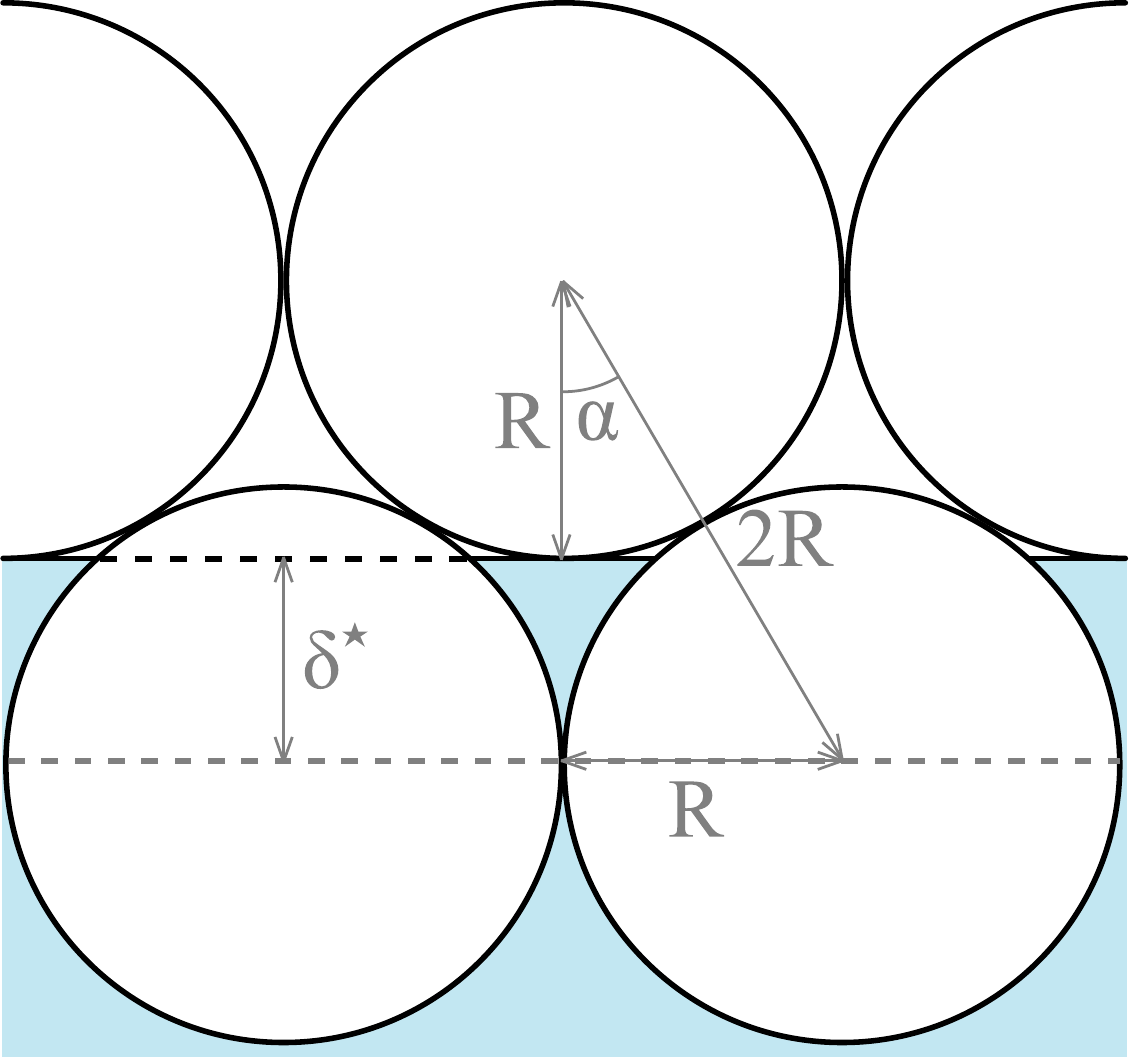}\label{fig:grains2}}
\hspace{.05\textwidth}
\subfigure[]{\includegraphics[width=0.4\textwidth]{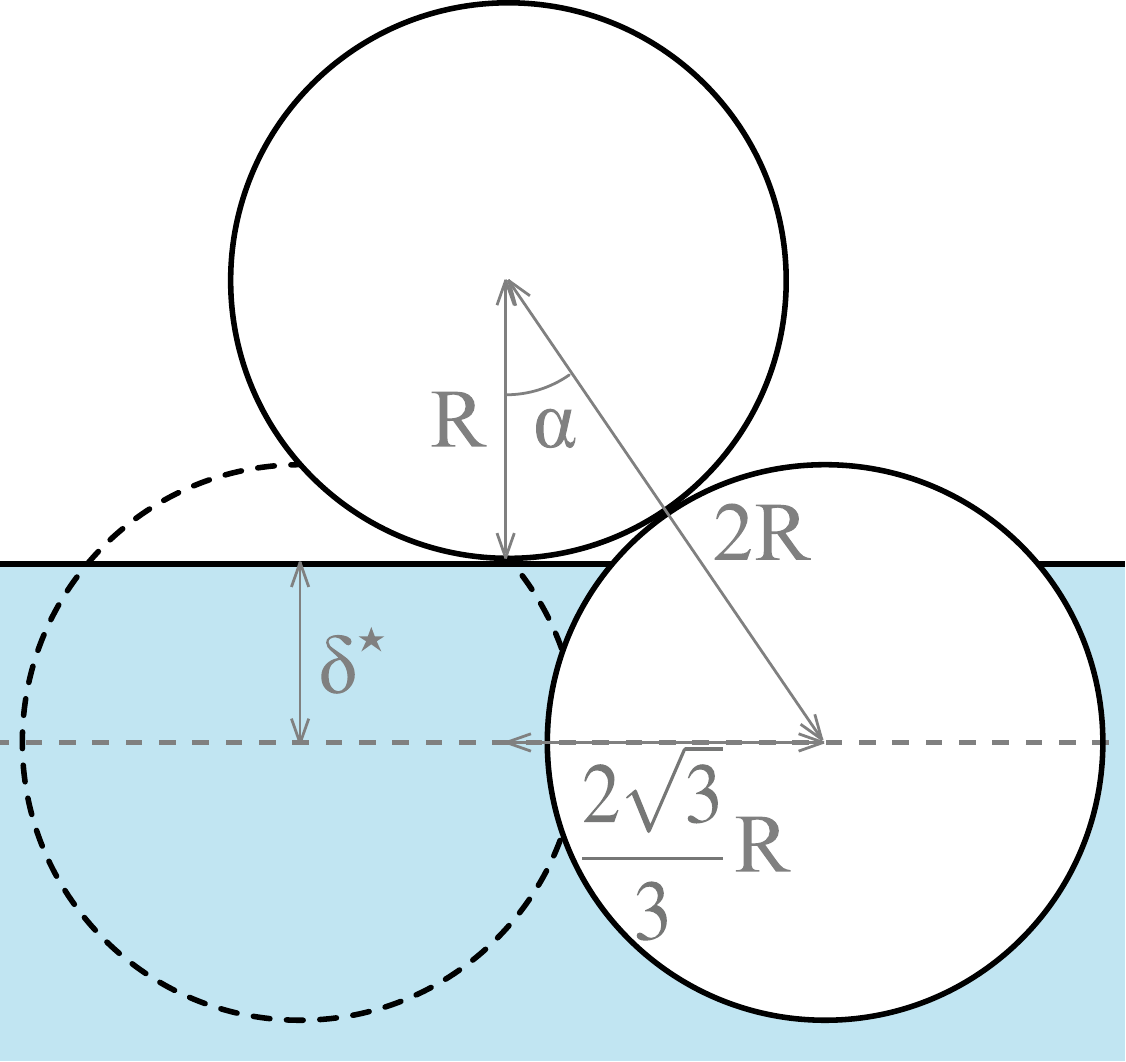}}
\caption{(a) Sphere at a curved interface. (b) Equilibrium state of a small bead at a liquid-gas interface. (c) Limit case for the impregnation in a 2-D close packing: the interface is tangent to the upper spheres. (d) Projection along the plane given by a median of the base of the tetrahedron and its summit for a 3-D pile. The dashed circle represents the two other spheres forming the base of the tetrahedron, out of the figure plane.}
\label{fig:grains}
\end{figure}

In order to explain these experiments, we first consider a single grain at the interface: since the spheres are denser than the liquid, gravity is balanced by interfacial forces. Let $\gamma$ be the surface tension of the liquid, $\psi$ the angle between the equatorial plane and the radius that connects to the contact line, $R_c$ the curvature of the interface and $\beta R_c$ the length of the liquid-air interface between the bead and the flat bath (Figure \ref{fig:grains}a). At small scale, the angle between the sphere and the interface must be $\theta$.  All these angles are linked by the geometrical relationship:
\begin{equation}
 \psi + \theta - \beta = \frac{\pi}{2}.
\label{eq:alphabeta}
\end{equation}


Keller \cite{keller1998surface} has shown that the vertical projection of the pressure forces is equal to the weight of the volume of liquid bounded by the horizontal free surface of the bath, 
the wetted surface of the body, and the vertical cylinder of radius $R \cos \psi$ and height $\xi=R_c (1- \cos \beta) $ (in white below the dashed line in Figure \ref{fig:grains}a). At equilibrium, the vertical projection of forces on the body can be written as a balance between the surface tension force $2 \pi R \gamma \cos \psi \sin \beta$ (corresponding to the weight of the volume in gray in Figure \ref{fig:grains}a) and an effective weight $\frac{4}{3}\pi R^3g \rho_{eff} $, gathering the buoyancy and the weight of the sphere. This leads to:

\begin{equation}
\sin \beta = \frac{2 R^2 g \rho_{eff} }{3 \gamma \cos \psi} =\frac{\rho_{eff}}{\rho}\frac{R^2}{a^2}\frac{2}{3\cos \psi},
\end{equation}
where $a=\sqrt{\frac{\gamma}{\rho g}}$ is the capillary length.
In our case, the radii are much smaller than the capillary length, and the ratio of densities is of order unity. Thus $\sin \beta$ is small, on the order of $10^{-2}$ for $R \sim 100~\mu m$. For $\theta >0$, we can neglect the effect of the weight and consider a flat interface as indeed observed in the experiments (insert in Figure \ref{fig:tvsVf}). 
This corresponds to $\psi+\theta=\frac{\pi}{2}$. The sphere equator stabilizes at a height $\delta$ below the flat interface (Figure \ref{fig:grains}b), where $\delta$ is given by:
\begin{equation}
 \delta= R \cos \theta.
\label{eq:delta1}
\end{equation}

\subsection{2-D pile}

Based on the experimental observation in Figure \ref{fig:setup}b, we consider a compact pile of grains (Figure \ref{fig:grains}c). As the equilibrium previously discussed is achieved for the first layer, the liquid may reach the second layer of grains. If it happens, the contact line is no longer at equilibrium and it will move up into the pile until it satisfies both a flat interface and a contact angle $\theta$ on the second layer. As it did for the first layer, it reaches the next spheres, and the same mechanism applies: no equilibrium is possible as long as there are dry grains. Liquid impregnates the pile, until the last monolayer of grains gets trapped at the interface. As the liquid rises, lower spheres become surrounded by liquid, and thus detach from the rest of the pile and fall into the bath, as seen in Figure \ref{fig:impr}c. Conversely, if the liquid does not reach the second layer of grains, only the first layer contacts the liquid and no grain detaches from the interface.

The limit between these two situations corresponds to $\theta=\theta^\star_{2D}$, for which the interface is tangent to the spheres of the second layer (Figure \ref{fig:grains}c). The distance between two equatorial planes is $R + \delta^\star$, which can also be written, introducing the angle $\alpha$ between the line linking the sphere centers from different layers with the normal to the first layer (Figure \ref{fig:grains}c):
\begin{equation}
R + \delta^\star = 2 R \cos \alpha.
\label{eq:delta2}
\end{equation}

In the compact 2-D case, the centers of monodisperse grains form equilateral triangles, which yields $\alpha = 30^\circ $. Together with Eq. (\ref{eq:delta1}) and (\ref{eq:delta2}), this provides the value of the critical contact angle $\theta^\star_{2D}$ for impregnating a 2-D pile (assembly of infinite cylinders):
\begin{equation}
\theta^\star_{2D}= \arccos ( \sqrt{3} -1) \approx 43^\circ.
\label{eq:theta2d}
\end{equation}

If the contact angle $\theta$ is larger than this critical value (and smaller than $90^\circ$), there is a local minimum of the surface energy of the system: impregnation can be blocked. The system may not reach its global minimum of energy, owing the existence of an energy barrier $\Delta E$ (per unit length) given by:
\begin{equation}
\frac{\Delta E}{2 \pi R \gamma} = \frac{\sin\theta - \sin \theta^\star + (\theta^\star -\theta )\cos \theta}{\pi},
\label{eq:deltaE}
\end{equation}
This expression becomes, at leading order in $\theta-\theta^\star$:
\begin{equation}
\frac{\Delta E}{2 \pi R \gamma} = \frac{ \sin \theta^\star}{2 \pi} (\theta-\theta^\star)^2.
\end{equation}
For a bead's diameter $2 R \sim 100\mu m$, the magnitude of the energy barrier typically is $2 R \Delta E  \sim 10^{-12}~\mathrm{J}$, much larger of course than thermal energy.

\subsection{3-D geometry}
Similar effects are expected in 3-D, and Shirtcliffe \textit{et al.} \cite{shirtcliffe2006critical} used surface free energies to predict the existence of a critical contact angle of wicking of $50.73^\circ$ for a compact pile. We can obtain this result using geometrical considerations. In a 3-dimensional compact pile of spheres, the grains form a tetrahedral network. The discussion above is still correct as well as Eq. (\ref{eq:delta1}) and (\ref{eq:delta2}), but the relative position of the successive layers is slightly modified, leading to a different value of $\alpha$, now given by $\sin \alpha = \frac{\sqrt{3}}{3}$ (Figure \ref{fig:grains}d). This provides the critical angle for impregnating a 3-D pile, also proposed by B\'an \textit{et al.}\cite{ban1987condition} and Shirtcliffe \textit{et al.}\cite{shirtcliffe2006critical}:

\begin{equation}
\theta^\star_0=\arccos (\sqrt{\frac{8}{3}} -1) \approx 51^\circ.
\label{eq:theta3d}
\end{equation}


In our experiments, contrasting with previous approaches \cite{ban1987condition,shirtcliffe2006critical}, we measure the contact angle directly on the grains, as discussed in section 1. Moreover, we bring the grains in contact with the liquid without confinement. Using this experimental protocol allows us to detect impregnation as it occurs over a single grain layer, improving the precision of the measurement.
The observed critical contact angles (Table \ref{tablo}) are indeed close to $\theta^\star_{0} \approx 51^\circ$, yet slightly larger (an additional deviation appears when $R$ is higher than $100~\mu m$, revealing the influence of gravity, neglected in the model). Different hypothesis can be proposed to explain this little (yet systematic) difference:
 

\begin{itemize}
\item Experimentally, wicking is reported as soon as the first grain detaches. Thus if the actual critical contact angle $\theta^\star$ locally differs from $\theta^\star_0$, our experimental criterion will only determine the highest possible critical contact angle.

\item In Figure \ref{fig:setup}c, one can see that the diameter of the grains is not perfectly fixed, to which correspond the standard deviations reported in Table \ref{tablo}. A small polydispersity can modify the geometry inside the pile, so that the local value of the critical angle $\theta^\star$ can be different from the monodisperse one, $\theta^\star_0$. In section 4, we investigate an elementary case of polydispersity, two-sized piles.

\item The interface was supposed to be flat, which is not the case if the difference of pressure through the interface is not negligible. In section 5, we discuss the effect of pressure for a pile of thickness comparable to the capillary length.

\item The determination of $\theta^\star_0$ assumes a close packing of spherical particles. As discussed in section 6, if the compacity of the pile is lower, or if the particle are slightly elongated, defects in the pile can appear, which modifies the wetting transition. 
\end{itemize}

\section{Polydispersity} \label{sect:poly}

\begin{figure}[h]
\centering
\includegraphics[width=0.4\textwidth]{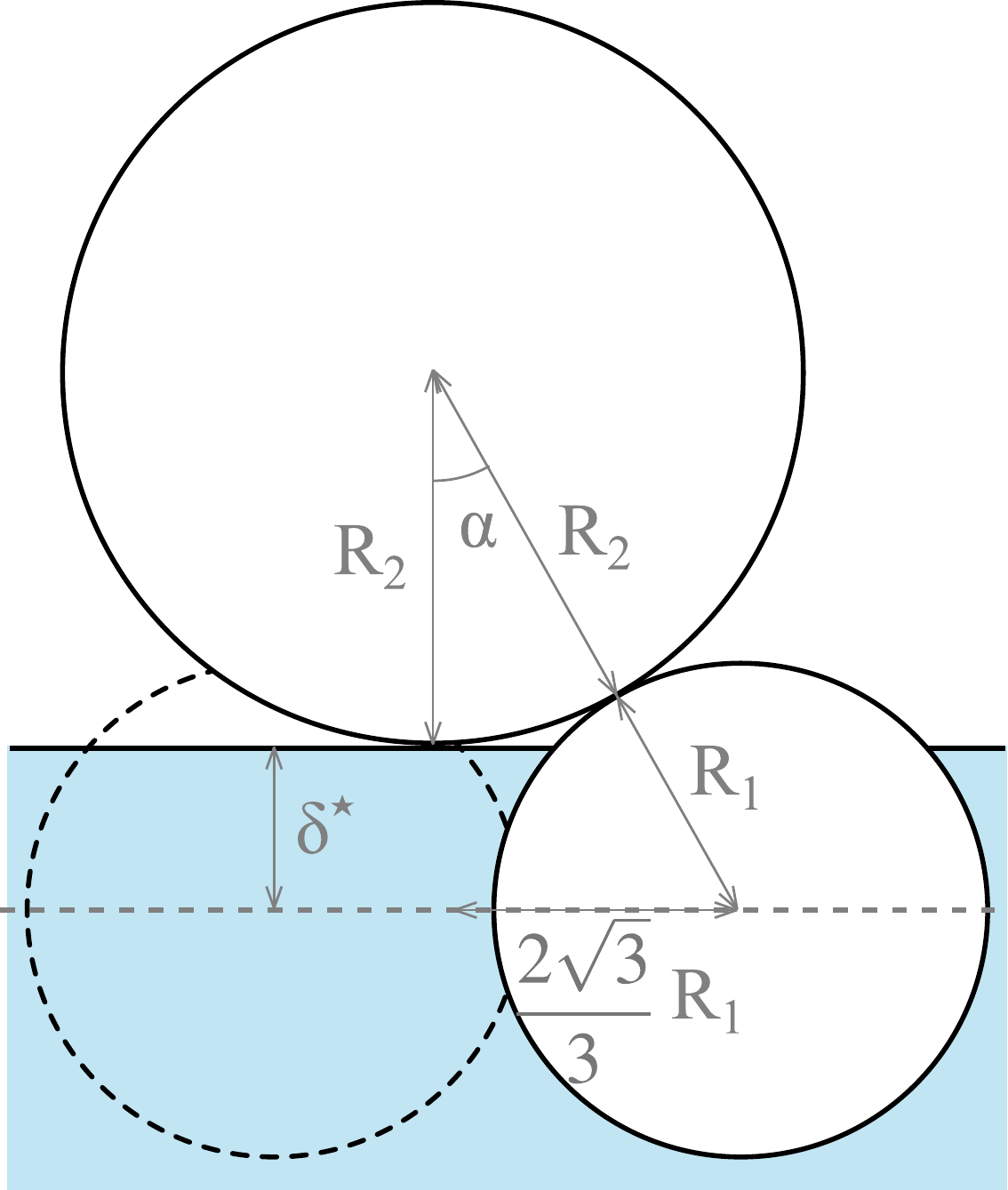}
\caption{Limit case of impregnation for a bilayer of bidisperse grains. The dashed circle represents the spheres of the lower layer out of the figure plane.}
\label{fig:poly2}
\end{figure}

As observed in Figure \ref{fig:hist}, there is a small dispersion of bead radii. B\'an \textit{et al.} considered a polydisperse packing and anticipated theoretically that impregnation should be easier ($\theta^\star$ larger) than in a monodisperse pile, although they found no difference experimentally \cite{ban1987condition}.

The theoretical determination of the critical angle is similar to the monodisperse case in 3 dimensions, but geometry is modified by the bidispersity of grains (Figure \ref{fig:poly2}). Eq. (\ref{eq:delta1}) can now be written for the lower layer:
\begin{equation}
\delta = R_1 \cos \theta.
\label{eq:poly1}
\end{equation}
In addition, we have:
\begin{equation}
R_2 + \delta = (R_1 + R_2 ) \cos \alpha,
\label{eq:poly2}
\end{equation}
where $\alpha$ is determined by:
\begin{equation}
\sin \alpha = \frac{2}{\sqrt{3}} \frac{R_1}{R_1 + R_2 }.
\label{eq:poly3}
\end{equation} 

Using equations (9), (10) and (11), we obtain the critical angle $\theta^\star$ as a function of the ratio $r=\frac{R_1}{R_2}$ ($<1$):

\begin{equation}
\cos \theta^\star=\frac{ \sqrt{1+2r -\frac{r^2}{3}} -1}{r}.
\label{eq:thetapoly}
\end{equation}

If the upper spheres are much larger than the lower ones ($r\rightarrow0$), the critical contact angle goes to $0^\circ$. Interestingly, this equation remains the same if upper beads are the small ones ($r>1$), and it is valid as long as upper spheres are large enough to stand upon the lower layer, that is, $r<3+2\sqrt{3}$. In the monodisperse limit ($r\rightarrow1$), Eq. (\ref{eq:thetapoly}) gives the same result than Eq. (\ref{eq:theta3d}). Additionally, for a small polydispersity, it yields at leading order in $(r-1)$:
\begin{equation}
\cos \theta^\star = \cos \theta^\star_0+(1-\frac{\sqrt{6}}{2})(r-1).
\end{equation}
Since $1-\frac{\sqrt{6}}{2}$ is negative, $\theta^\star$ increases with $r$ and it exceeds $\theta^\star_0$ for small spheres on big ones ($r>1$). This situation is the one favorable to wicking and thus should determine the path followed during the wicking of a random bidisperse pile. That said, for applications where wicking has to be guaranteed or prevented, Eq. (\ref{eq:thetapoly}) allows identifying the most stringent contact angle condition, which has to be met in the most unfavorable case of a local segregation.

  \begin{figure}[t]
\centering
\subfigure[]{\includegraphics[width=0.32\textwidth]{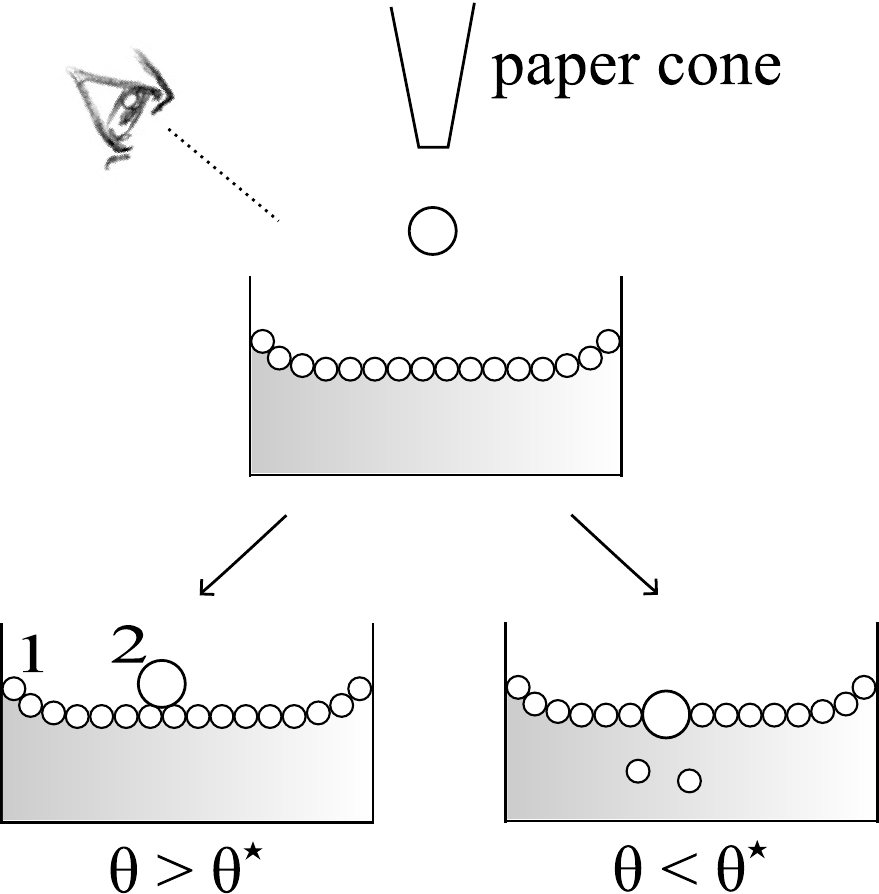}\label{fig:poly-schema}}
\subfigure[$v_f=0.80,\theta=48^\circ \pm5^\circ$]{\includegraphics[width=0.32\textwidth]{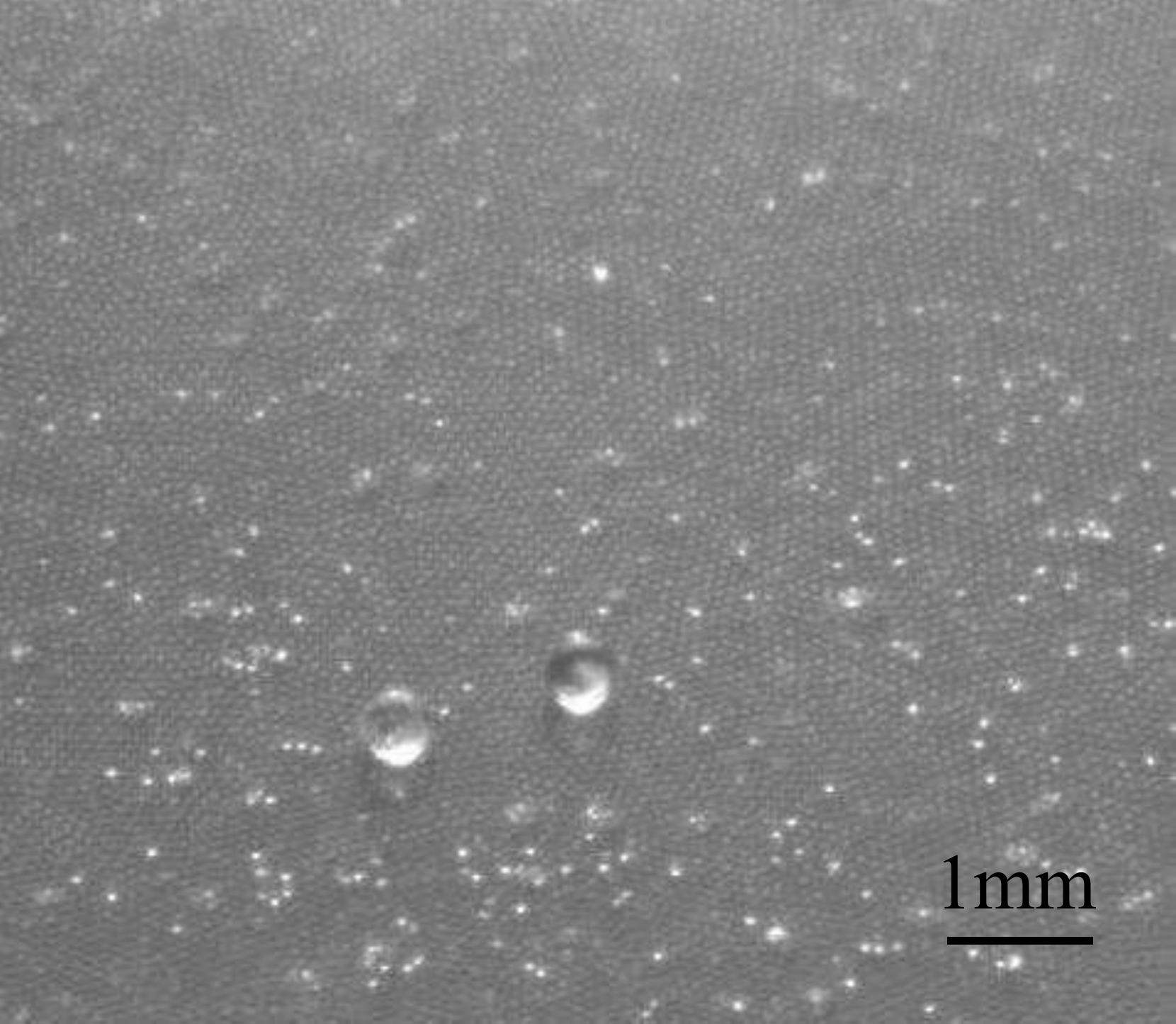}\label{fig:poly-ni}}
\subfigure[$v_f=0.97,\theta=36^\circ \pm5^\circ$]{\includegraphics[width=0.32\textwidth]{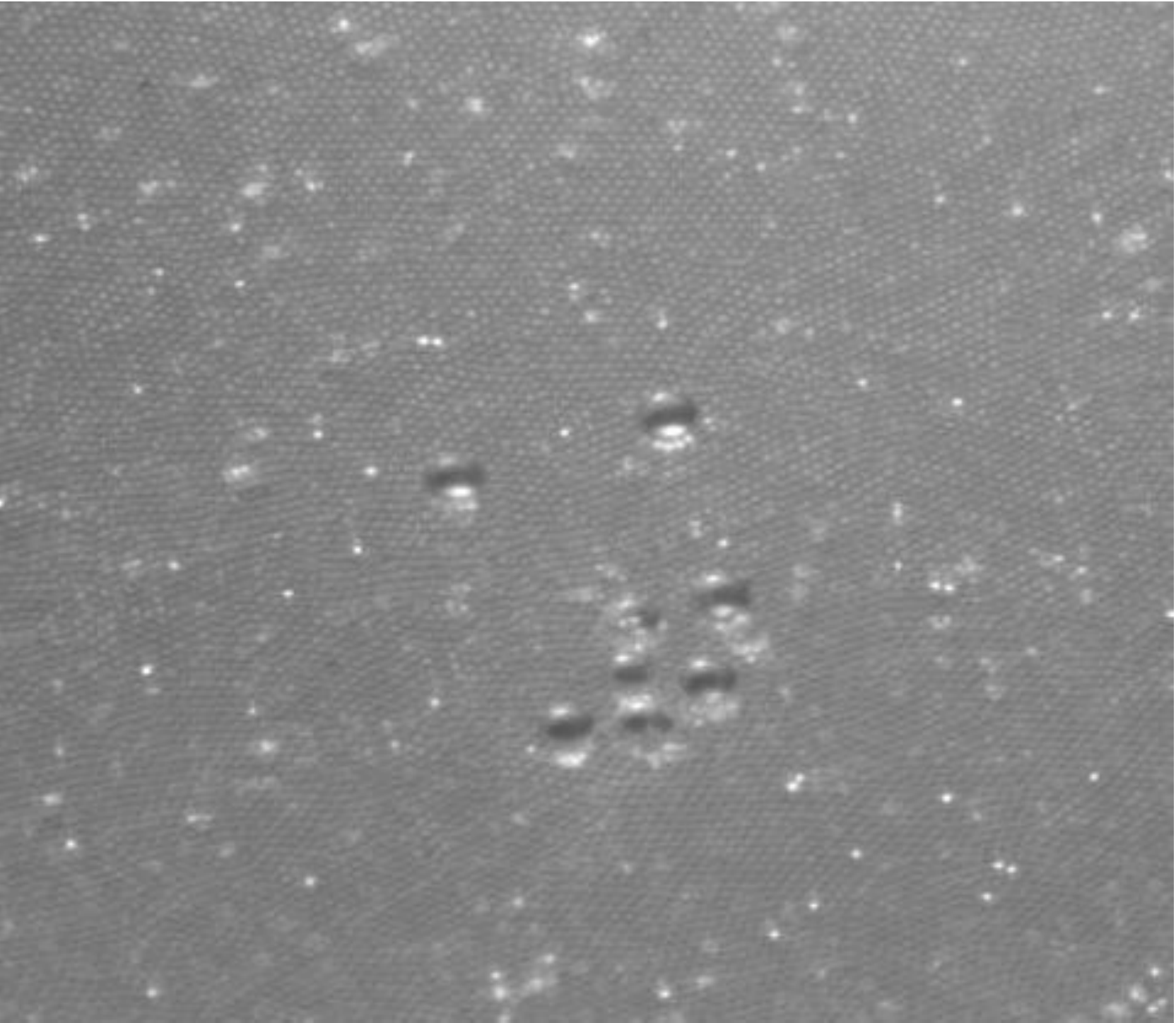}\label{fig:poly-i}}
\caption{(a) Set-up of the bidisperse experiment. (b)  $R_2=256\pm13~\mu m$ beads stay on top of the base monolayer of radius $R_1=52\pm2~\mu m$. (c) Same experiment, with a lower contact angle ($\theta<\theta^\star$). The $R_2$ spheres partially pass through the monolayer, and only their top poles remain visible. }
\label{fig:poly}
\end{figure}

\begin{figure}[t]
\centering
\includegraphics[width=0.6\textwidth]{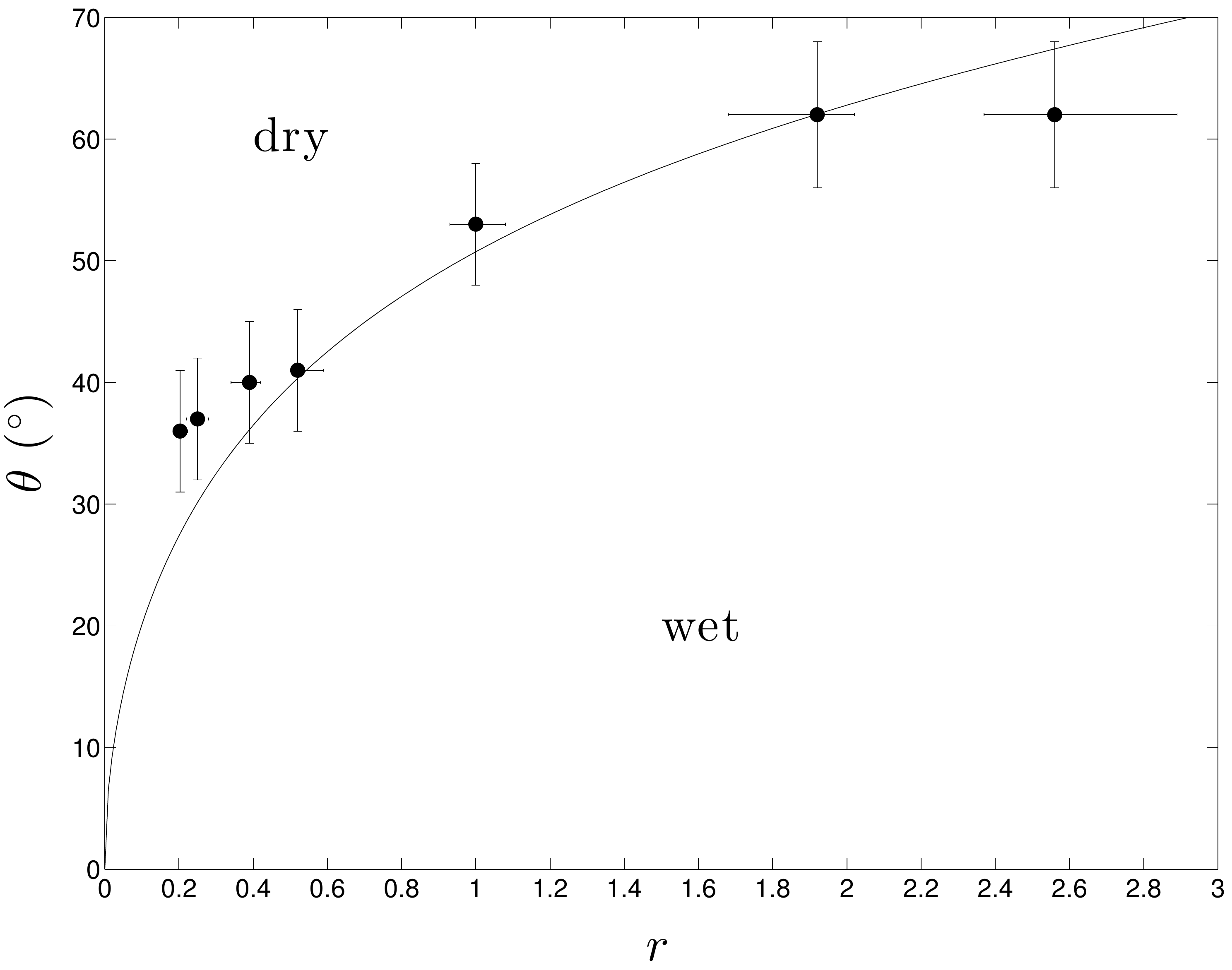}
\caption{Wicking phase diagram for a bidisperse powder, in terms of contact angle $\theta$ versus the ratio of grain radii $r=\frac{R_1}{R_2}$. Dots are data of the critical angle $\theta^\star$, and the solid line is Eq. (\ref{eq:thetapoly}). It is found that the smaller $r$ is, the easier the wicking.}
\label{fig:polydth}
\end{figure}
 In order to investigate experimentally the effect of polydispersity, we achieved two layers systems, each one being formed of beads of a given size. Combinations of spheres presented in Table \ref{tablo} were used to obtain values of $r$ between $0.2$ and $2.5$.
A monolayer of beads of radius $R_1$ is first placed at the surface of the bath, with a contact angle $\theta$. Then beads of radius $R_2$ are added one by one (Figure \ref{fig:poly}a). We observe two regimes, depending on $\theta$. If $\theta>\theta^\star$, the beads of radius $R_2$ stay upon the monolayer of beads of radius $R_1$, and no impregnation is observed (Figure \ref{fig:poly}b). In contrast, for $\theta<\theta^\star$, the beads of radius $R_2$ are wetted and pass partially through the interface, between the beads of the monolayer (Figure \ref{fig:poly}c). Some $R_1$-beads detach from the interface and sink into the liquid while only the north pole of the upper sphere remains dry. The measured critical angle $\theta^\star$ is highly sensitive to the polydispersity of these elementary piles since it varies from $\theta^\star=36^\circ$ for $r=0.2$ to $\theta^\star = 62^\circ$ for $r=2$, as reported in Figure \ref{fig:polydth}. Eq. (\ref{eq:thetapoly}) is coherent with the data, even if the value of the critical contact angle seems slightly underestimated (Figure \ref{fig:polydth}). This discrepancy can be due to defects in the packing of the base monolayer, which is investigated in section 6. However, this experiment emphasizes again how critical geometry is in the wicking of grains.


Coming back to the "monodisperse" experiments of section 1, the size of beads slightly varies from a layer to another, hence modifying the local critical contact angle. The dispersion of radius typically yields $r \approx 1.1$ (Table \ref{tablo}), which generates a critical angle $\theta^\star$ of $52^\circ$: polydispersity can explain part of the difference between the experimental data in Table \ref{tablo} and the theoretical value of $\theta^\star_0$ expected from Eq. (\ref{eq:theta3d}).

\section{Hydrostatic forcing} \label{sect:forcing}

\begin{figure}[h]
\centering
\subfigure[$h<h^\star$]{\includegraphics[width=0.3\textwidth]{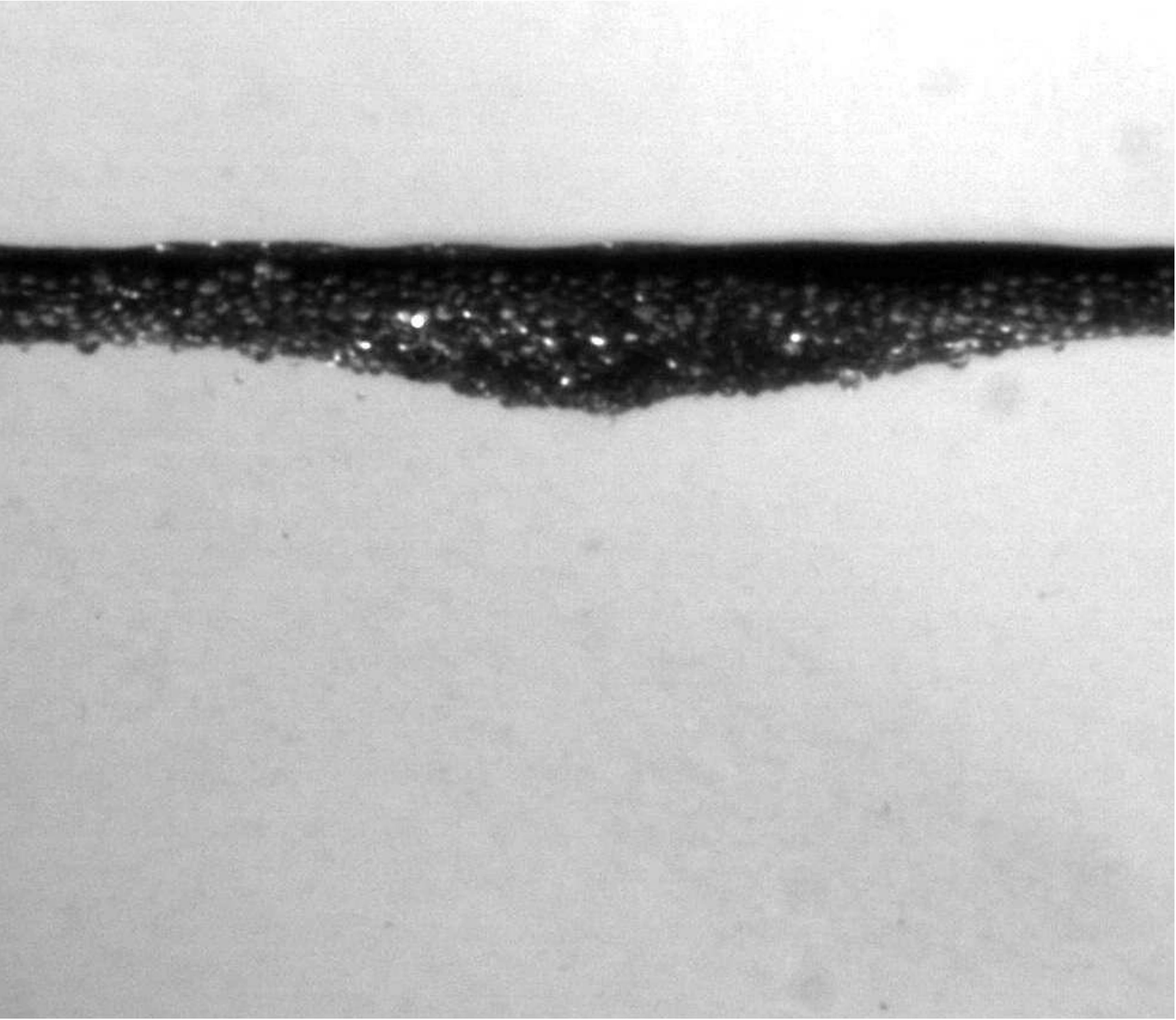}\label{fig:Hini}}
\subfigure[$h=h^\star$]{\includegraphics[width=0.3\textwidth]{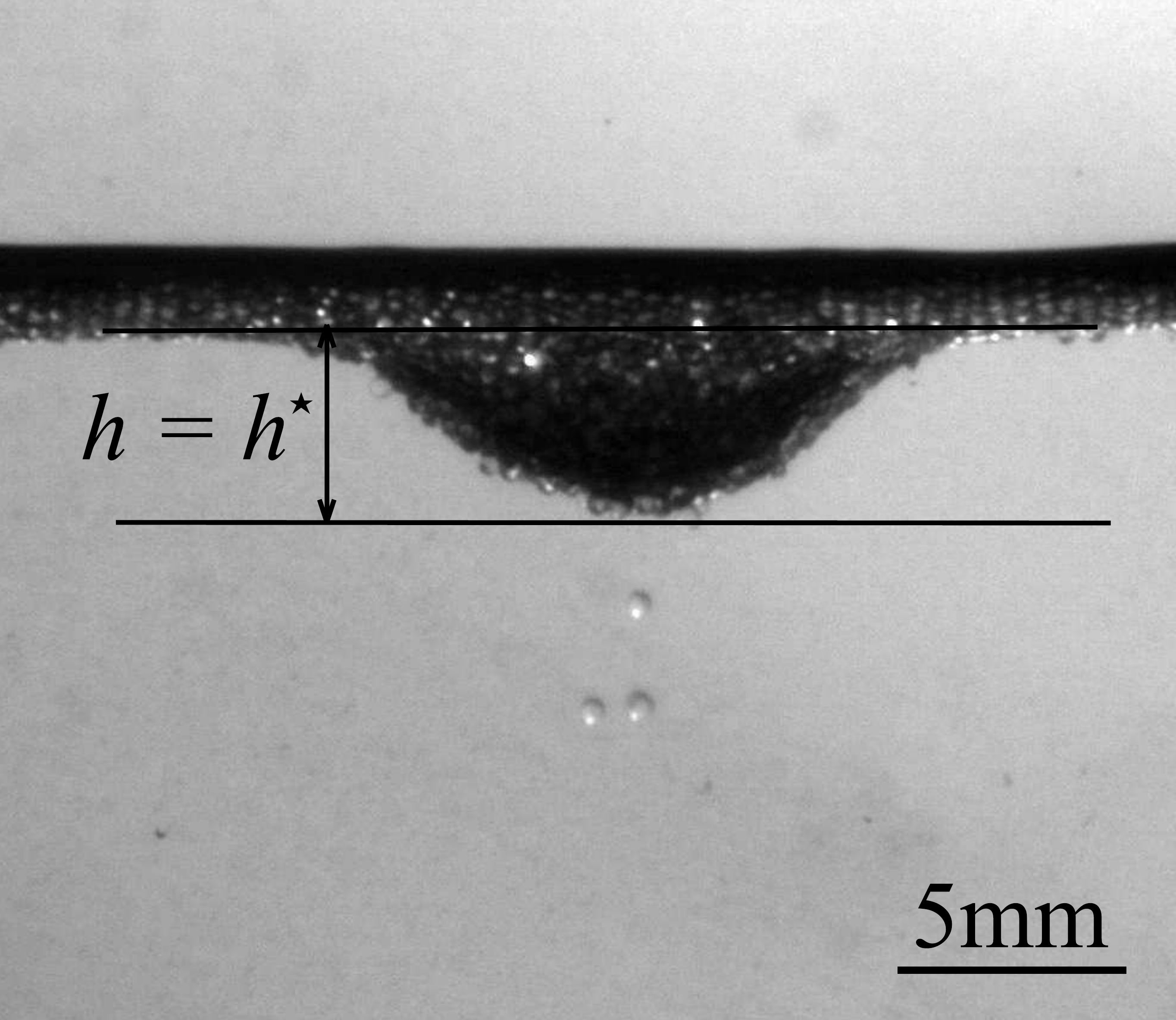}\label{fig:Hlim}}
\subfigure[$h>h^\star$]{\includegraphics[width=0.3\textwidth]{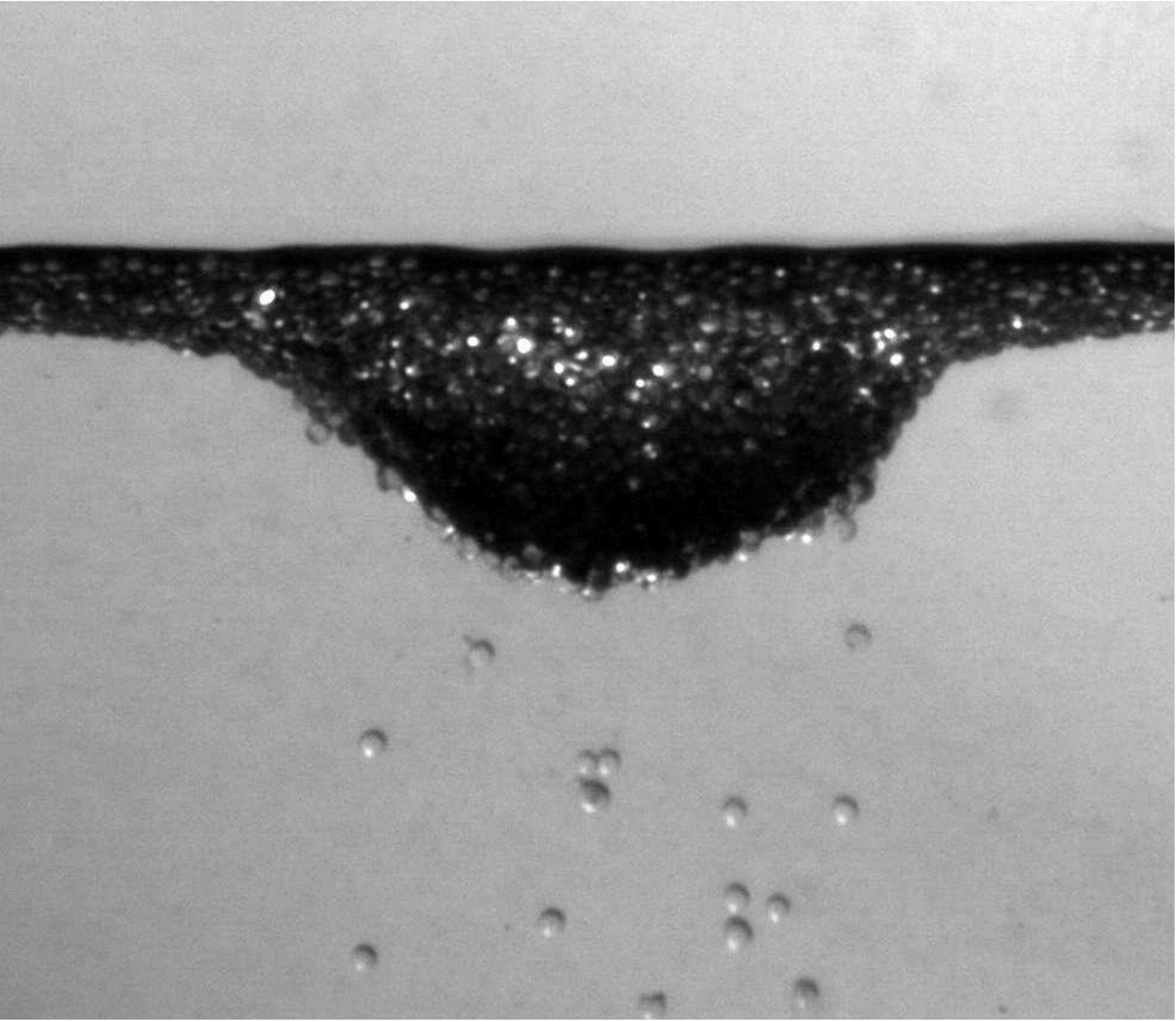}\label{fig:Himp}}
\caption{Forced impregnation at an ethanol volume fraction $v_f=0.33$, varying the pile thickness $h$. The contact angle here is $\theta = 80^\circ \pm 5^\circ > \theta^\star_0$. When $h$ is bigger than $h^\star$, impregnation is observed.}
\label{fig:hydroforcing}
\end{figure}

\begin{figure}[t]
\centering
\includegraphics[width=0.8\textwidth]{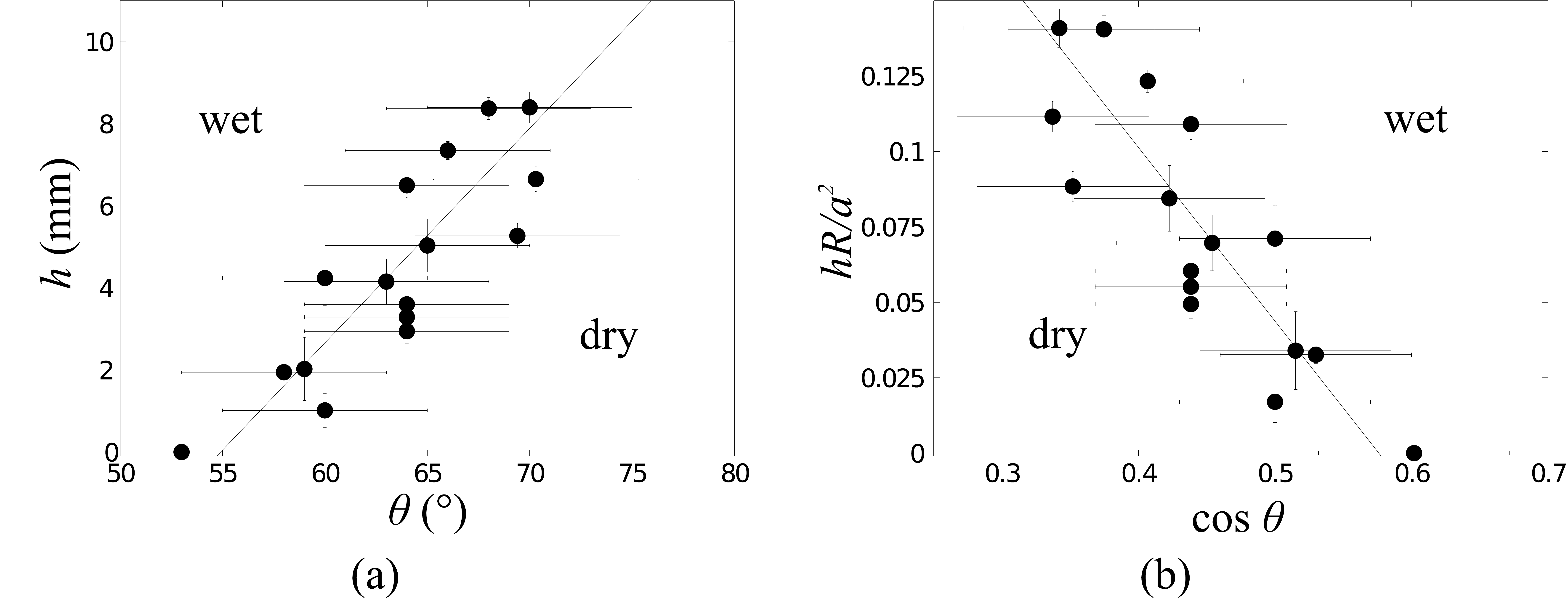}
\caption{(a) Phase diagram for the thickness $h$ of the pile, as a function of the contact angle $\theta$. The dots correspond to the limit between wet and dry states, and thus represent the critical thickness $h^\star$ of wicking for beads of radius $R=52~\mu m$. The line is a linear regression of slope $30~\mathrm{mm/rad}$. (b) Phase diagram in terms of normalized thickness of powder versus the cosine of the contact angle, and critical values for beads of radius $R=52~\mu m$.} 
\label{fig:hvst}
\end{figure}

Even if the thickness $h$ of the dry pile was set as small as possible in section 2, spheres can be coarse enough to make $h$ comparable to the capillary length $a$. In this case, hydrostatic pressure plays a role, changing the shape of the interface between beads and thus the limit for impregnation. The experiment in Figure \ref{fig:hydroforcing} shows the influence of the pile thickness on wicking. On a liquid with a contact angle $\theta>\theta^\star_0$ (no wicking), we locally feed the surface with beads to increase the thickness. For $h$ lower than a threshold $h^\star$, the pile remains dry (Figure \ref{fig:hydroforcing}a). As the thickness $h$ reaches $h^\star$, a few grains detach from the interface (Figure \ref{fig:hydroforcing}b). For thicker piles, the dry powder is impregnated, and many beads fall in the bath (Figure \ref{fig:hydroforcing}c). The reference $h=0$ is taken at the lowest point of the meniscus in the transparent cell, before any grain is added. Figure \ref{fig:hvst}a shows a phase diagram $(\theta,h)$ separating dry and wet states. At a given height $h$, wicking only occurs if $\theta$ is smaller than a critical value $\theta^\star$, which increases with $h$. The two domains are separated by a line, of slope $30~\mathrm{mm/rad}$  for $R=52~\mu m$ .



\begin{figure}[t]
\centering
\subfigure[]{\includegraphics[width=0.4\textwidth]{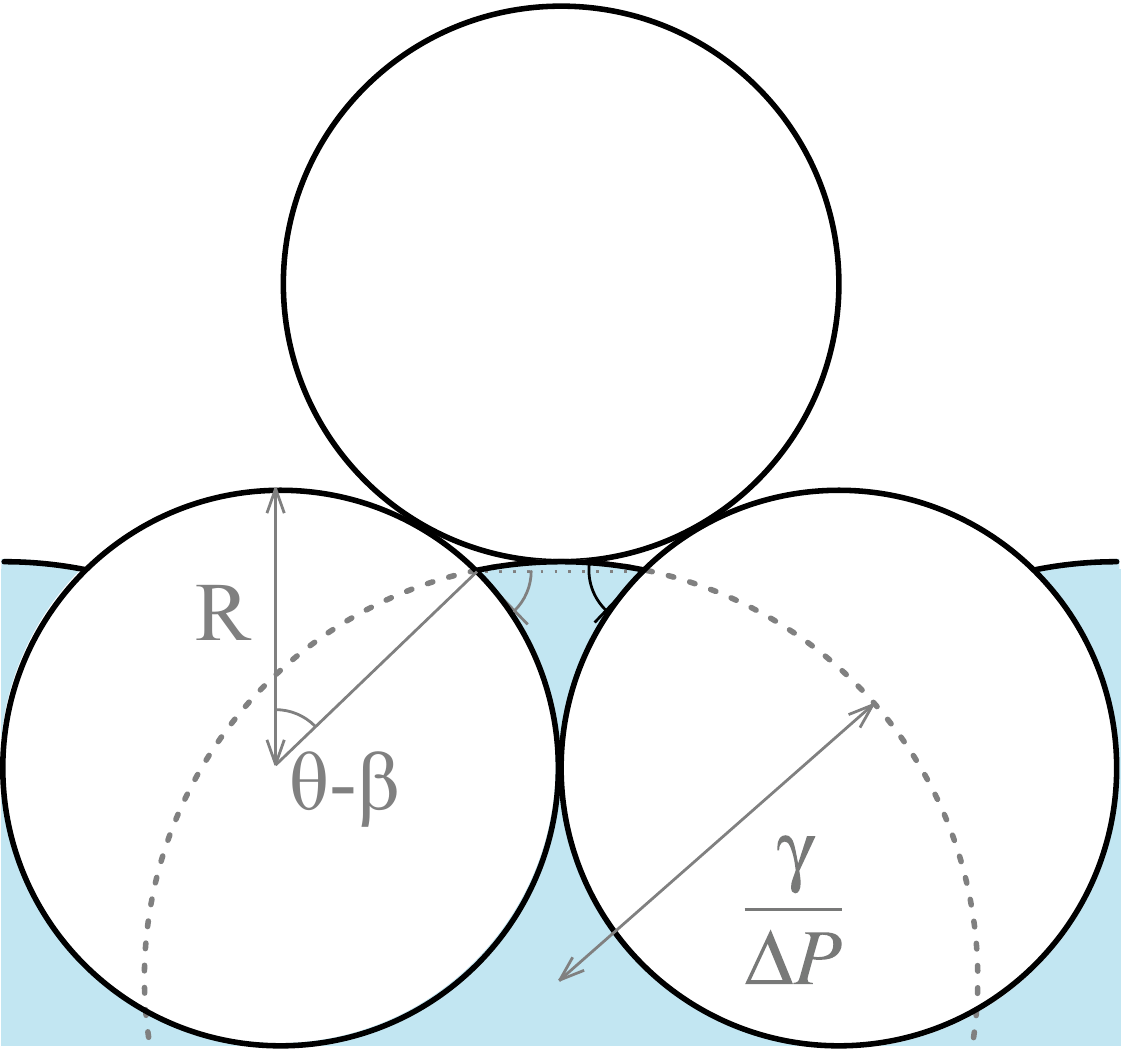}}
\hspace{.05\textwidth}
\subfigure[]{\includegraphics[width=0.4\textwidth]{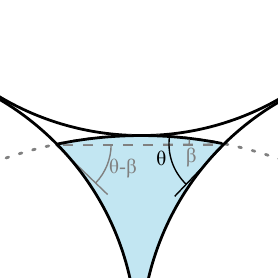}}

\caption{Limit of impregnation with forcing, and close up on the meniscus. Due to the curvature of the meniscus, the liquid reaches the second layer of grains while no impregnation would have been observed with a flat meniscus and the same contact angle. }
\label{fig:forcing}
\end{figure}

If the grain radius is comparable to $a$, or if there is a pressure difference $\Delta P$ between the liquid and the gas, the interface gets curved between the beads, as illustrated in Figure \ref{fig:forcing}. The associated Laplace pressure compensates the pressure $\Delta P$, and changes the condition for impregnation. The Laplace equation gives the radius of curvature of the meniscus at equilibrium $\frac{ \gamma}{\Delta P}$, where $\gamma$ is the surface tension of the liquid. The sign of this curvature is related to the sign of the forcing: a larger pressure in the liquid helps the impregnation process, elevating the highest point of the meniscus (Figure \ref{fig:forcing}). The expression for $\delta$ changes, while Eq. (\ref{eq:delta2}) remains correct. If $\beta$ is the angle between the meniscus at the contact point with the sphere and the equatorial plane (Figure \ref{fig:forcing}), one gets:
\begin{equation}
\delta = R \cos( \theta - \beta) +\frac{ \gamma}{\Delta P} ( 1 - \cos \beta ).
\label{eq:forcing1}
\end{equation}

The distance between the two contact points can be written as a function of either $\beta$ or $\theta-\beta$ (Figure \ref{fig:forcing}), which leads to:
\begin{equation}
\frac{ \gamma}{\Delta P} \sin \beta = R ( 1 - \sin (\theta - \beta) ).
\label{eq:forcing2}
\end{equation}

To determine $\beta$, we can consider a situation close to the flat meniscus. Assuming $\beta \ll 1$, and a meniscus radius larger than the grain diameter ($\frac{R \Delta P}{\gamma} \ll 1$), Eq. (\ref{eq:forcing2}) simplifies at the leading order into:
\begin{equation}
\beta =\frac{R \Delta P}{ \gamma} (1 - \sin \theta).
\label{eq:betaforcing}
\end{equation}
Hence Eq. (\ref{eq:forcing1}) becomes:
\begin{equation}
\delta = R \cos \theta + \frac{R^2 \Delta P}{2 \gamma} \cos^2 \theta.
\label{eq:deltaforcing}
\end{equation}
%
%

As shown in section 3, the critical angle for impregnation with a flat meniscus is given by $  \cos \theta^\star_0 = \delta / R$. With a curved meniscus, the critical contact angle $\theta^\star$ is the angle $\theta$ solving Eq. (\ref{eq:deltaforcing}), which yields at leading order:
\begin{equation}
\cos \theta^\star = \cos \theta^\star_0 - \frac{R \Delta P}{2 \gamma} \cos^2 \theta^\star_0.
\label{eq:thetaf}
\end{equation}

If the pressure $\Delta P$ is negative, which means a higher pressure in the air, it is harder to impregnate the pile ($\theta^\star<\theta^\star_0$): the liquid has to wet more the surface to wet the grains. Conversely, if the pressure is positive, a liquid with $\theta$ larger than $\theta^\star_0$ can invade the pile. If the grain radius is comparable to the capillary length, the curvature is set by a balance between the weight of the sphere, buoyancy and surface tension. Its sign will depend on the balance of the first two forces: if the spheres are denser than the liquid, the curvature is positive, and piles of such spheres can be impregnated with a contact angle higher than $\theta^\star_0$, as indeed observed with the largest beads in Table \ref{tablo}.\\

As seen in Figure \ref{fig:hydroforcing}, the pressure can be imposed hydrostatically: we have $\Delta P= \rho g h$, where $h$ is the depth of the deepest point of the pile (Figure \ref{fig:hydroforcing}b). According to Eq. (\ref{eq:thetaf}), the cosine of the critical contact angle should be linear in $h$.
A liquid with a contact angle $\theta>\theta^\star_0$ invades the powder if $h$ is higher than a critical value $h^\star$ given by:
\begin{equation}
h^\star= \frac{2 a^2}{R} \frac{\cos \theta^\star_0 - \cos \theta}{\cos^2 \theta^\star_0},
\label{eq:hf}
\end{equation}

where $\cos \theta^\star_0 = \sqrt{\frac{8}{3}}-1$ (Eq. (\ref{eq:theta3d})) and $a^2\sim3.1\pm 0.1 \mathrm{mm^2}$ (a quantity almost constant in the tested range of ethanol volume fraction). In Figure \ref{fig:hvst}b, the normalized critical height $h^\star$ is observed to decrease as increasing $\cos \theta$, as predicted by Eq. (\ref{eq:hf}). However, the experimental value of the slope is $-0.6$, one order of magnitude smaller than the slope predicted by our model, $\frac{-2}{\cos^2 \theta^\star_0} \approx -5$. This discrepancy might be due to large errors on both $h^\star$ and contact angle, measured in a narrow domain of $\cos \theta$.

\section{Eccentricity and defects in the pile} \label{sect:defect}

\begin{figure}[h]
\centering
\subfigure[$\theta=105^\circ$, $R=52~\mu m$]{\includegraphics[width=0.3\textwidth]{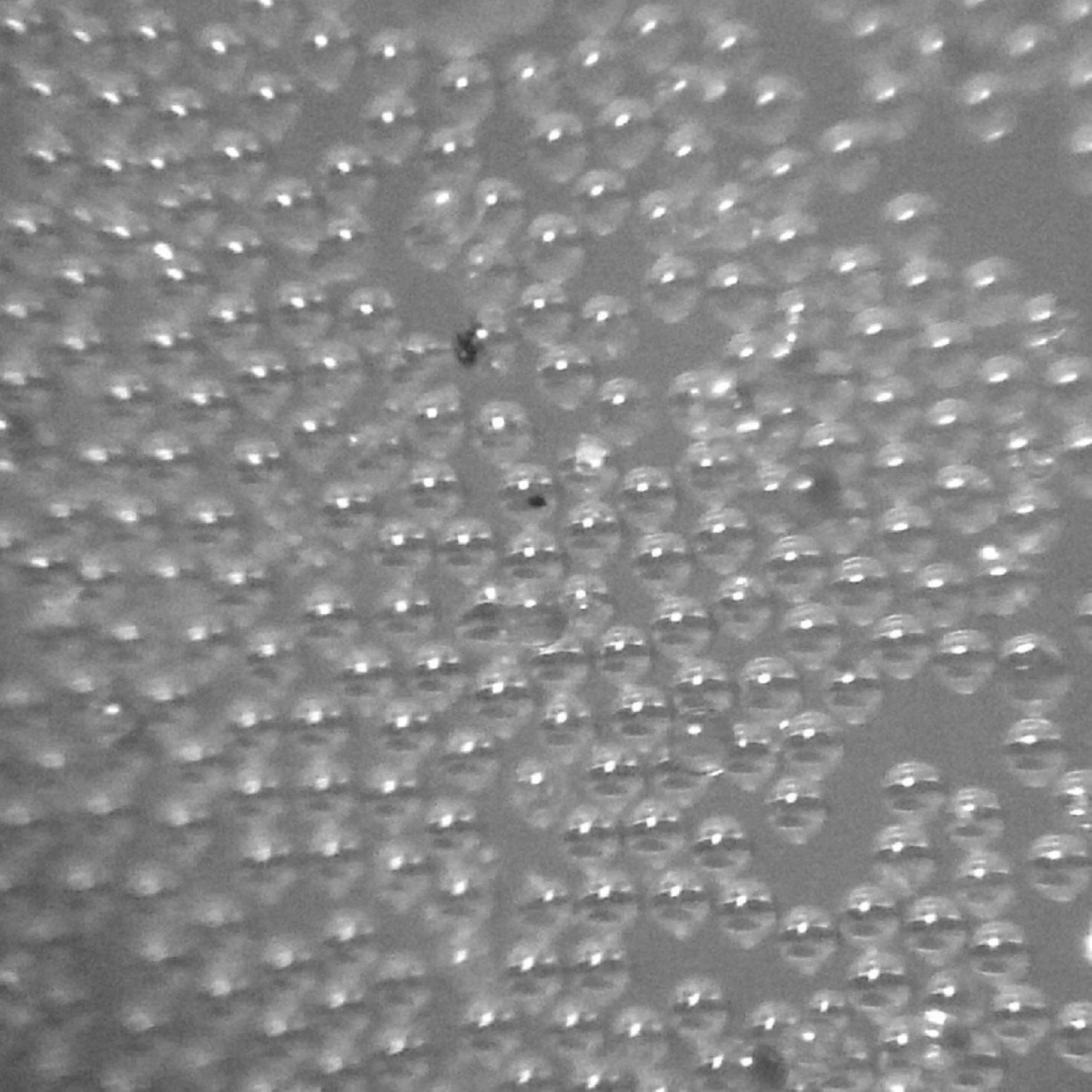}\label{fig:compa100-0}}
\subfigure[$\theta=105^\circ$, $R=100~\mu m$]{\includegraphics[width=0.3\textwidth]{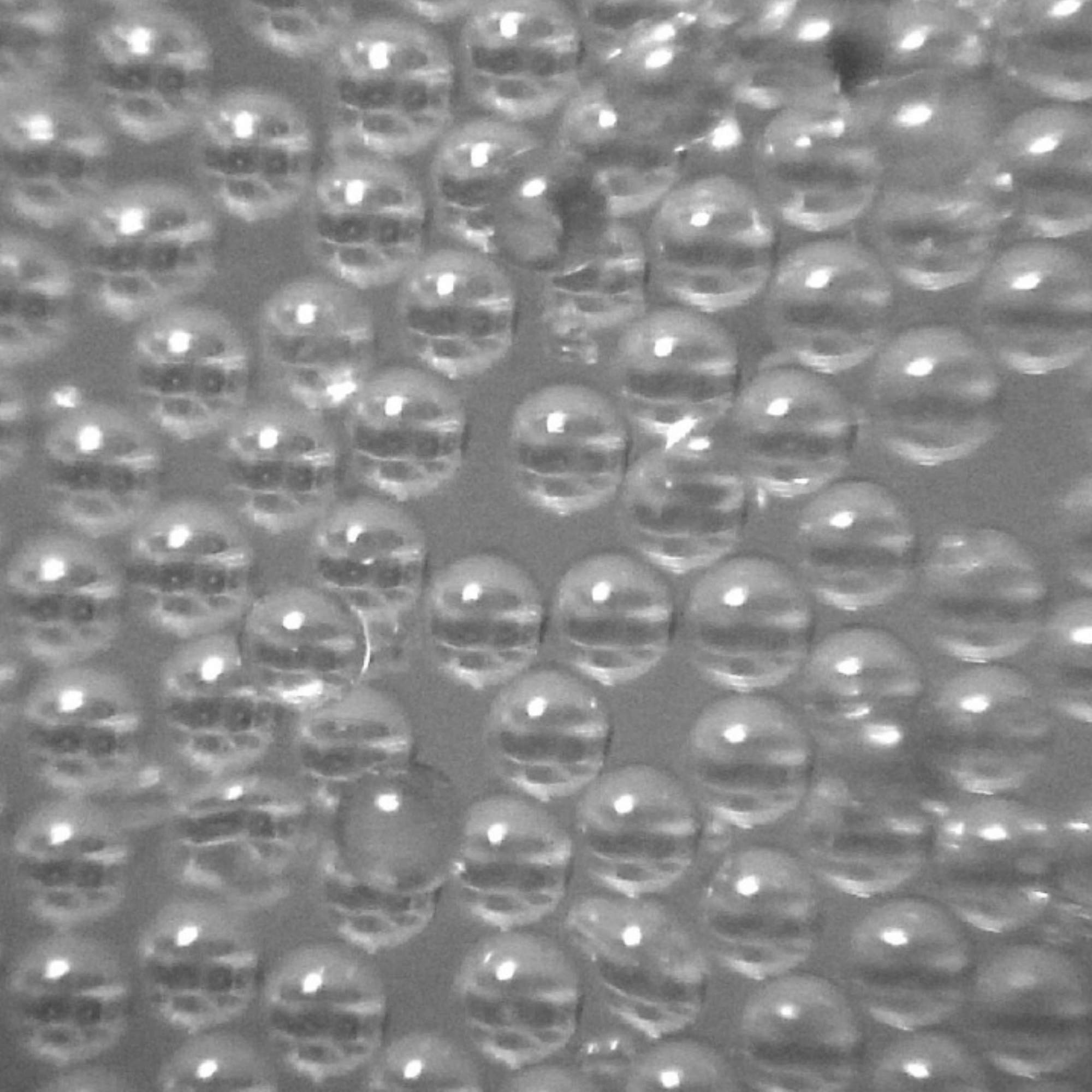}\label{fig:compa200-0}}\\
\subfigure[$\theta=45^\circ$, $R=52~\mu m$]{\includegraphics[width=0.3\textwidth]{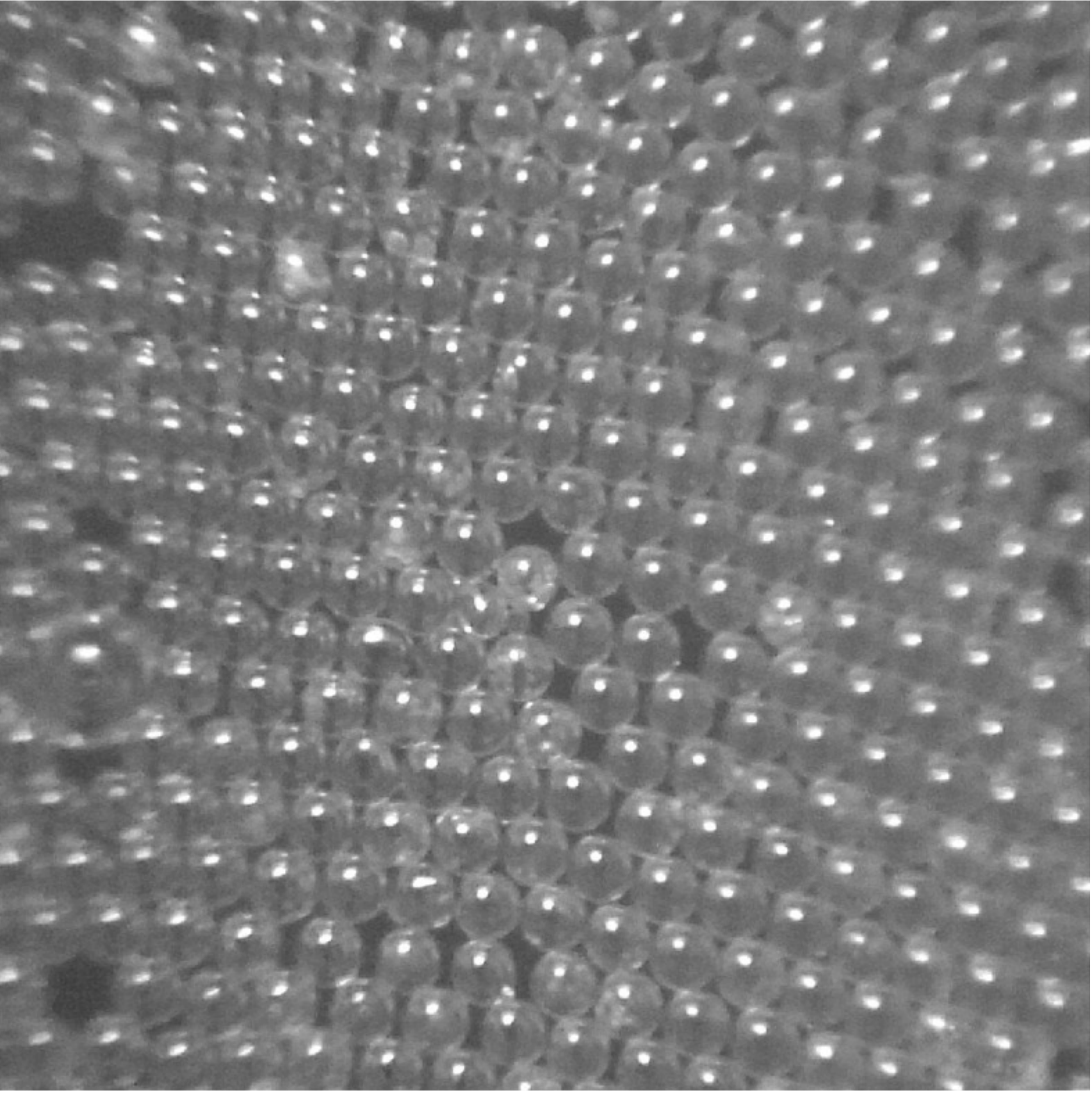}\label{fig:compa100-80}}
\subfigure[$\theta=45^\circ$, $R=100~\mu m$]{\includegraphics[width=0.3\textwidth]{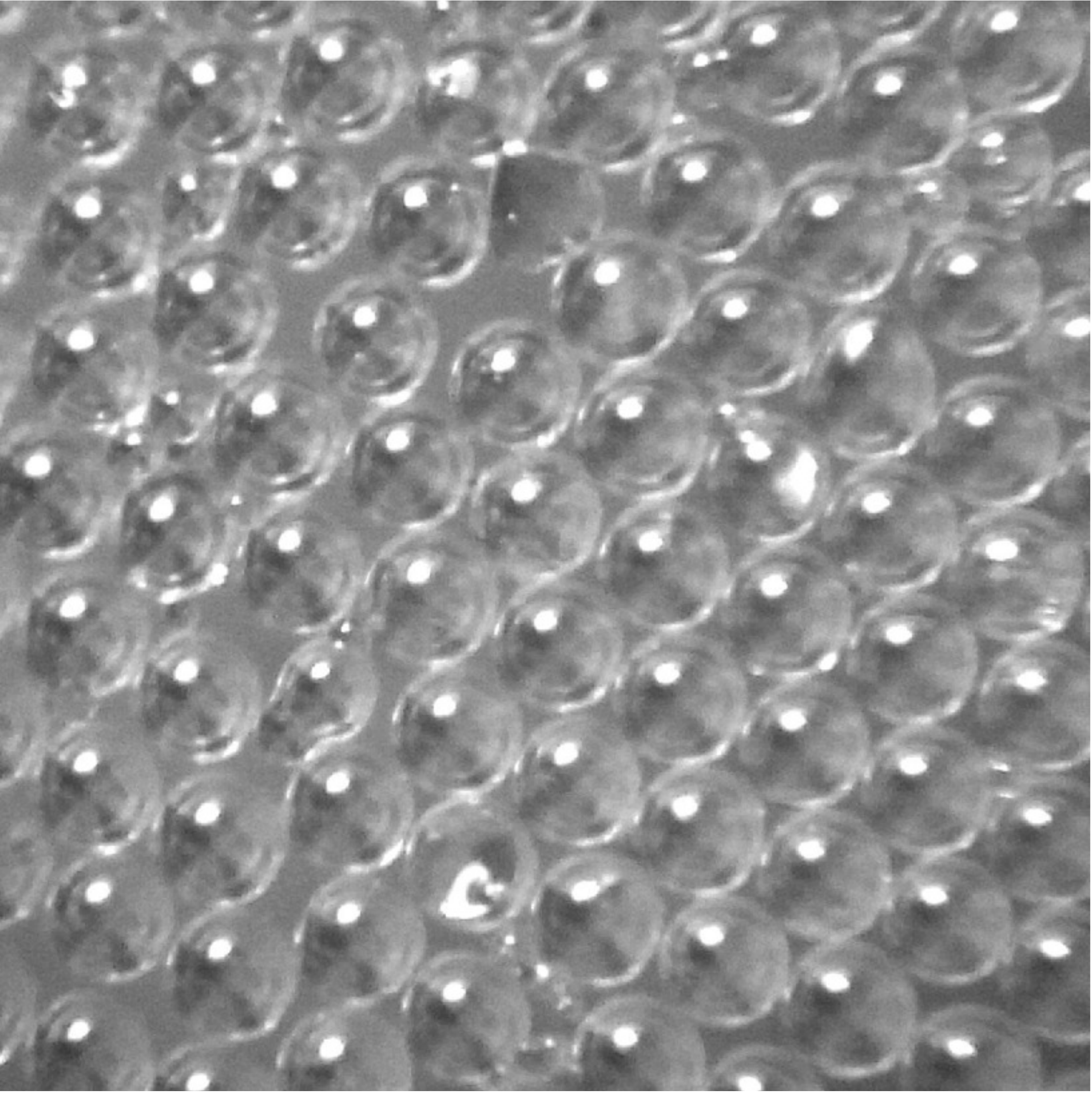}\label{fig:compa200-80}}

\caption{Compacity of monolayers of grains of radius $R$ and contact angle $\theta$ at the surface of water/ethanol solutions. Pictures are taken from above through a binocular x20.}
\label{fig:compa}
\end{figure}

\begin{figure}[t]
\centering
\subfigure[]{\includegraphics[width=0.3\textwidth]{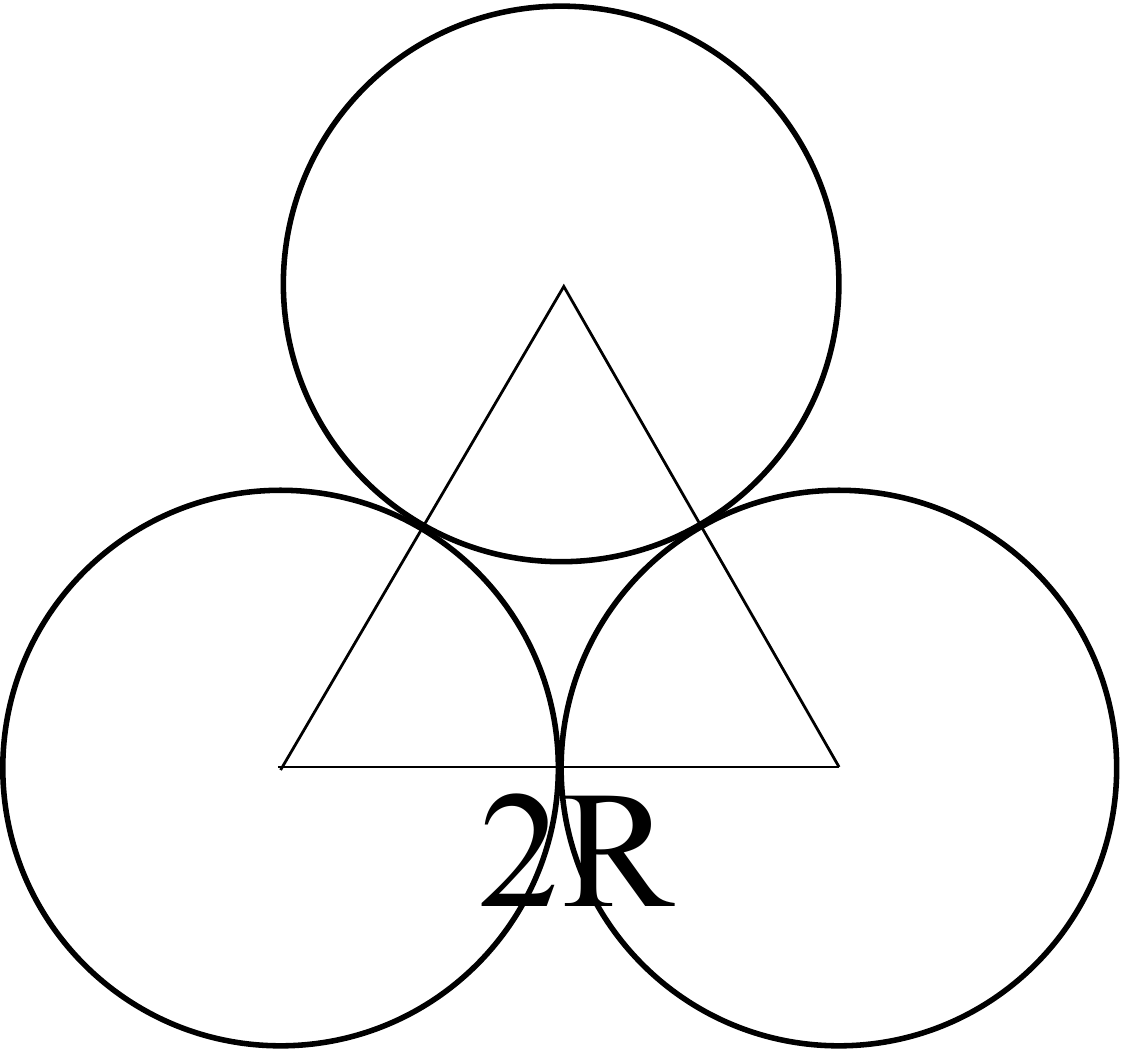}\label{fig:compact1}}
\subfigure[]{\includegraphics[width=0.3\textwidth]{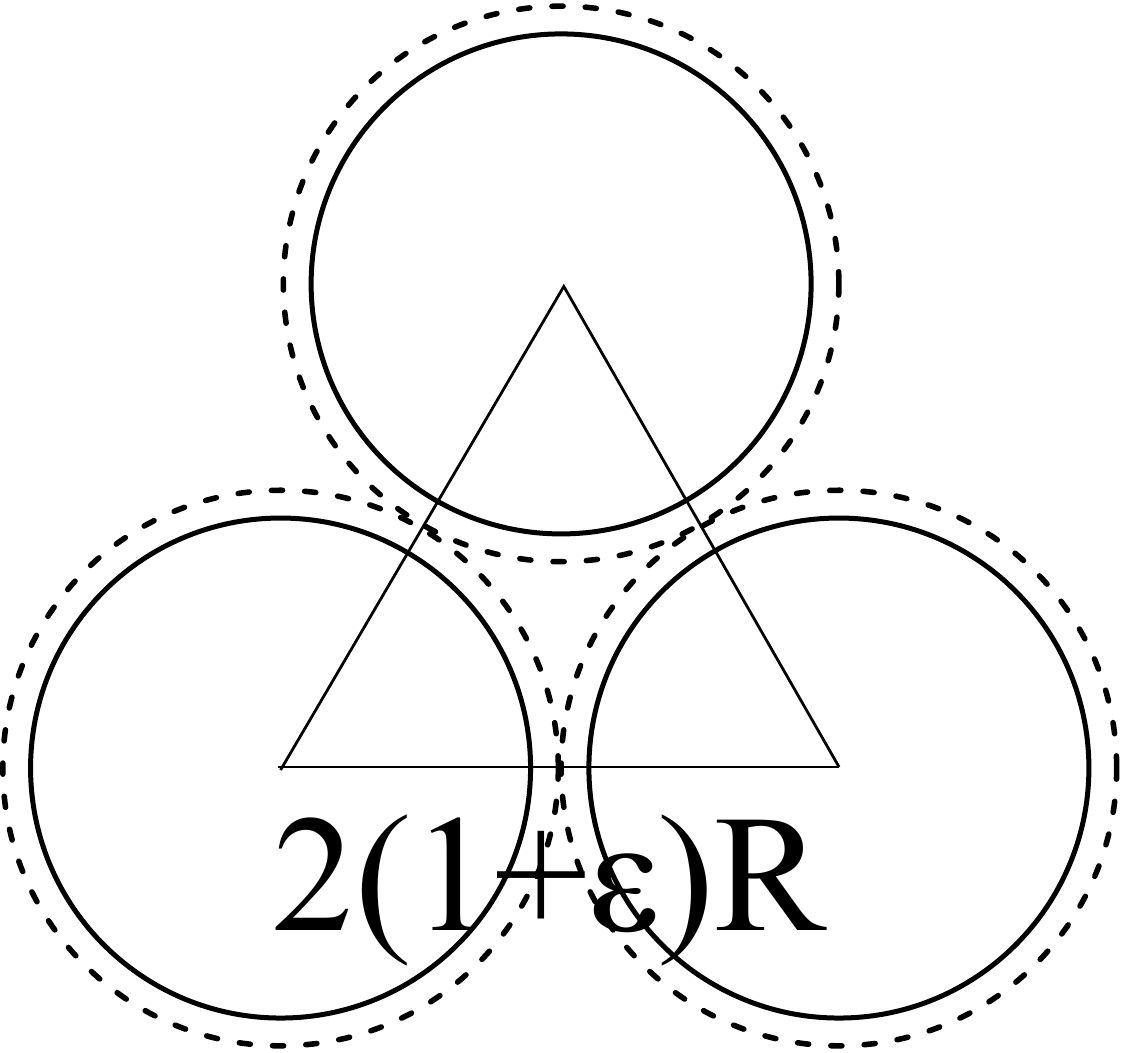}\label{fig:compact2}}
\subfigure[]{\includegraphics[width=0.35\textwidth]{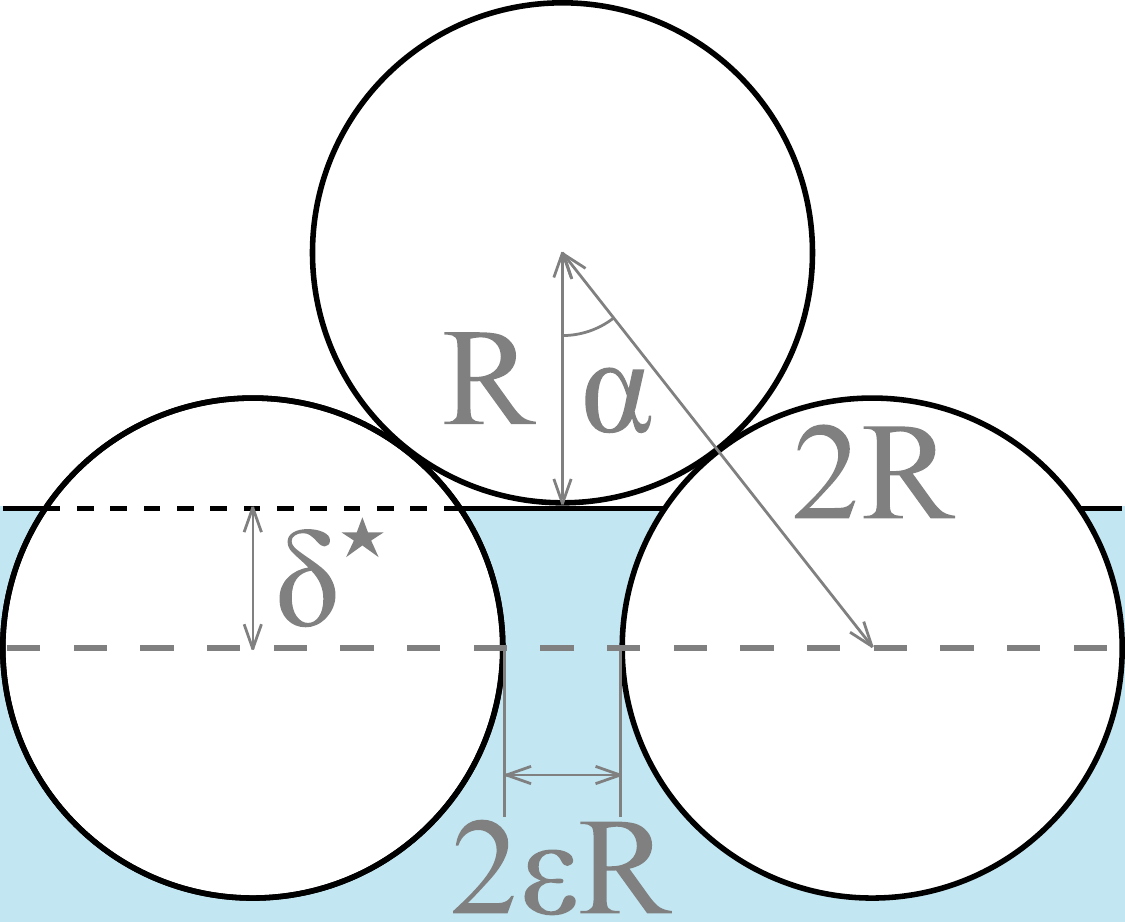}}
\caption{Surface fraction occupied by spheres at the liquid/air interface: in the triangle, the surface of the grains is $\pi  R^2/2$. In the close packing case (a), the total surface is $\sqrt{3}R^2$ while it is $\sqrt{3} (1+\epsilon)^2 R^2$ if there is a gap $2 \epsilon R$ between spheres (b). (c): As model defects in the first layer of the pile, we consider gaps $2 \epsilon R$ between the grains. }
\label{fig:compact}
\end{figure}

Defects in the packing of the pile may also affect impregnation. Indeed, the observation of a monolayer of beads (Figure \ref{fig:compa}) shows deviations from close packing. Packing depends on the way the monolayer is prepared, and defects especially occur if the liquid does not impregnate a monodisperse pile ($\theta>\theta^\star$). In this section, we consider as a model defect gaps between spheres. Typical values for the gap are obtained by comparing the surface fraction occupied by the spheres to a compact situation. In a 2-D close packing, the surface fraction occupied by spheres is $\Phi_{c}=\frac{\pi}{2 \sqrt{3}}$ (Figure \ref{fig:compact1}). If the spheres have a gap of $2 \epsilon R$ between them, the compaction decreases to $\Phi=\frac{\Phi_c}{(1+ \epsilon)^2}$ (Figure \ref{fig:compact2}), so the dimensionless gap can be expressed as:
\begin{equation}
\epsilon = \sqrt{\frac{\Phi_c}{\Phi}}-1.
\end{equation}

From pictures such as shown in Figure \ref{fig:compa}, we can extract the compacity $\Phi$ and thus deduce a mean value for $\epsilon$, as reported in Table \ref{tablo:epsilon}. 
Experimentally, the monolayer has less defects when made on a wicking liquid for a monodisperse pile: a small excess of grains will be removed by the wicking of the pile, as described earlier (section \ref{sect:expmt}). In the polydisperse case and for $r>1$, we are in the opposite situation and thus expect the base monolayer to be poorly packed.

We focus on the situation favoring wicking, in particular when a defect is present in the first layer of grains (Figure \ref{fig:compact}c).
For a monodisperse pile, Eq. (\ref{eq:delta1}) and (\ref{eq:delta2}) are still correct. In the 3-D case, since the base of the tetrahedron is loose, as sketched in Figure \ref{fig:compact}b, $\alpha$ is given by $\sin \alpha = \frac{\sqrt{3}}{3}(1+\epsilon)$.
These equations lead to a modified critical contact angle, function of $\epsilon$:
\begin{equation}
\cos \theta^\star = \sqrt{\frac{8}{3} (1-\epsilon-\frac{\epsilon^2}{2})}-1,
\end{equation}
which becomes, for low $\epsilon$:
\begin{equation}
\cos \theta^\star= \cos \theta^\star_0-\sqrt{\frac{2}{3}}\epsilon.
\label{eq:thetaepsilon}
\end{equation}

These equations show that introducing defects in the packing ($\epsilon >0$) tends to increase $\theta^\star$, so that wicking becomes less demanding in term of contact angle - thus approaching the classical assumption of $\theta^\star=90^\circ$.
Applying Eq. (\ref{eq:thetaepsilon}) to the experimental values of \ref{tablo} leads to estimations of $\epsilon$ in the monodisperse case: we find $\epsilon = 4\%$ for $R=25~\mu m$ and $R=52~\mu m$, $\epsilon=7\%$ for $R=100~\mu m$ and $\epsilon=12\%$ for $R=256~\mu m$.  These results are coherent with the fact that smaller spheres form more compact layers, due to a relatively stronger interaction of the meniscus (Figure \ref{fig:compa100-80} and Figure \ref{fig:compa200-80}) \cite{mansfield1997equilibrium}. These estimations are however slightly higher than observed on monolayers (Table \ref{tablo:epsilon}), but the latter values are average quantitites, whereas the experimental protocol is sensitive to local larger values of $\epsilon$.

\renewcommand{\arraystretch}{1.8}
\begin{table}[t]
\centering
\begin{tabular}{|c|c|c|}
\hline
\hspace{0.05\textwidth}$R\: [\mu m]$ \hspace{0.05\textwidth} &  \hspace{0.05\textwidth}$\theta=45^\circ < \theta^\star$  \hspace{0.05\textwidth} & \hspace{0.05\textwidth}$\theta=105^\circ > \theta^\star$\hspace{0.05\textwidth}  \\
\hline
$52 \pm 2$ & $\Phi = 0.90$, $\epsilon = 0.4\%$ & $\Phi = 0.72$, $\epsilon = 12\%$ \\
\hline
 $100 \pm 5$ & $\Phi=0.87$, $\epsilon=2.1\%$ & $\Phi = 0.76$, $\epsilon = 9.2\%$ \\
\hline
$256 \pm 13$& $\Phi = 0.86$, $\epsilon= 2.7\%$ & $\Phi = 0.82$, $\epsilon=  5.2\%$ \\
\hline
\end{tabular}
\caption{\label{tablo:epsilon} Estimated values of the surface fraction of beads $\Phi$ and average relative gap $\epsilon$, corresponding to the different situations and radius.}
\end{table}


\section{Conclusion}
In order to study the wicking in granular media, we carried out experiments on piles of glass beads at the surface of a bath, and shown the existence of a critical angle $\theta^\star$ below which wicking occurs. This angle is significantly smaller than the one observed in capillary tubes ($90^\circ$), and its value is close to the value expected from models, around $51^\circ$. This critical angle can be modified by a pressure gradient across the liquid-air interface, defects in the pile compacity, or by polydispersity of the grains. These effects can be used in industrial processes to either help or prevent the wicking of a powder, depending on the field of application. Moreover, this study emphasizes the crucial role of geometry in the wicking of ordered powders. More generally, the geometry of porous media allows one to control the penetration or to prevent the invasion of this medium by a given liquid, which permits a wetting liquid to be repelled from a solid surface, as  observed with super-oleophobic materials \cite{tuteja2007designing}.



%

\end{document}